\documentclass[10pt]{article}
\usepackage{amsmath}
\usepackage{cite}
\usepackage{hyperref}
\usepackage{relsize}
\usepackage{amssymb}
\usepackage{graphics}
\usepackage{epsfig}
\usepackage{epstopdf}
\usepackage{verbatim}
\usepackage{color}
\usepackage{subcaption}
\usepackage{multirow}
\usepackage{textcomp}
\usepackage{float}
\usepackage{mathtools}
\usepackage[title]{appendix}
\usepackage{enumitem}

\begin{document}
	\title{ On minimal realization of Topological Lorentz Structures with one-loop Seesaw extensions in A$_4$ Modular Symmetry}
	\author {Monal Kashav\thanks{Electronic address:  monalkashav@gmail.com} and Surender Verma\thanks{Electronic address: s\_7verma@hpcu.ac.in}}	
	\date{\textit{Department of Physics and Astronomical Science,\\Central University of Himachal Pradesh, Dharamshala 176215, INDIA.}}
	\maketitle
	%\date{\today}
	\begin{abstract}
	The topological classification of one-loop Weinberg operator at dimension-5 leads to systematic categorization of one-loop neutrino mass models. All one-loop neutrino mass models must fall in one of these categories. Among these topological categories, loop extension of canonical seesaw scenarios is interesting in light of the current LHC run. Apart from one-loop contribution, these extensions result in dominant tree-level contribution to neutrino masses. The immediate remedy to obtain dominant one-loop contribution requires combination of flavor symmetries and enlarged field content. Alternatively, in this work, we propose a minimal way of realizing the topological structures with dominant one-loop contribution using modular variant of the permutation symmetries. In such a realization, no new fields are needed apart from those permitted by the topology itself. For the first time, we have realized one such topological Lorentz structure(T4-2-$i$) pertaining to one-loop extension of Type-II seesaw using modular A$_4$ symmetry. Here, modular weights play an important role in suppressing tree-level terms and stabilizing the particles running in the loop($N_i$, $\rho$ and $\phi$), thus, making them suitable dark matter candidates. In this work, we have explored the possibility of fermionic dark matter candidate where right-handed neutrino ($N_1$) is assumed to be lightest. We have, also, analyzed the compatibility of the model with neutrino oscillation data and obtained model predictions for effective Majorana mass $M_{ee}$ and $CP$ violation. Furthermore, the predictions on relic density of dark matter and its direct detection considering bound on lepton flavor violating process, $\mu\rightarrow e\gamma$ have, also, been investigated.
	\end{abstract}
	\maketitle
	\section{Introduction}
The Weinberg operator at the lowest dimension, $d=5$, allows lepton number violation by two units ($\Delta L=2$) and explains the smallness of neutrino masses \cite{Weinberg:1979sa}. UV completion of the Weinberg operator leads to well-known Type-I \cite{Minkowski:1977sc,Mohapatra:1979ia}, Type-II \cite{Magg:1980ut, Schechter:1980gr,Wetterich:1981bx} and Type-III \cite{Foot:1988aq} seesaw paradigms. In conventional Type I/II/III seesaw mechanisms, new heavy degree of freedom suppresses the neutrino mass. These heavy degrees of freedom are not accessible at current Large Hadron Collider (LHC) runs. Apart from the introduction of new degrees of freedom, fine-tuning of Yukawa couplings is required to correctly explain the small neutrino masses. Besides these seesaw scenarios, a more promising framework is radiative generation of neutrino mass at one-loop level as it explains non-zero neutrino mass and dark matter, simultaneously. Furthermore, some of LHC accessible new physics variants for neutrino mass generation are inverse \cite{Mohapatra:1986bd} or linear seesaw \cite{Akhmedov:1995ip,Akhmedov:1995vm} mechanisms at tree level, radiative mass generation through loop integrals \cite{Cai:2017jrq}, SUSY models with R-parity violation \cite{Aulakh:1982yn,Ellis:1984gi,Abada:2001zh} to name a few.

\noindent In general, Weinberg operator at one-loop \cite{Bonnet:2012kz} and two-loop \cite{AristizabalSierra:2014wal} leads to systematic topological classification of neutrino mass models. In fact, Weinberg operator leads to six topologies  Ti (i=1,2,...,6) of one-loop diagrams with four external legs \cite{Bonnet:2012kz}. Out of these six topologies T2 is discarded on the basis of dimensional disagreements. The topologies T3, T5 and T6 have one Lorentz structure whereas T1, T4 can have different topological structures depending on whether the fields running in the loop are of scalar or fermionic nature,
\begin{eqnarray}
&&\text{T1:\hspace{1.0cm} T1-$i$; T1-$ii$; T1-$iii$}, \nonumber\\
&&\text{T4:\hspace{1.0cm} T4-1-$i$; T4-1-$ii$; T4-2-$i$; T4-2-$ii$; T4-3-$i$; T4-3-$ii$.} \nonumber
\end{eqnarray}
\noindent All the Lorentz structures corresponding to these six topologies can be grouped in three categories \textit{viz.}, (i) divergent one-loop extensions of seesaw: T4-1-$i$, T4-2-$ii$, T4-3-$ii$, T5, T6  (ii) finite one-loop diagrams giving leading contribution to non-zero neutrino mass naturally: T1-$i$, T1-$ii$, T1-$iii$, T3 and (iii) finite one-loop extension of seesaw: T4-1-$ii$, T4-2-$i$, T4-3-$i$. For the topological Lorentz structures of T1 and T3 (category (ii)), it is always possible to have one-loop dominant contribution to non-zero neutrino mass by forbidding the tree-level contribution using discrete or U(1) symmetry \cite{Bonnet:2012kz}.  Further, on the basis of symmetry arguments it can be shown(discussed in Section 3.1) that T4 topology(category (iii)) carries dominant tree-level  contribution in addition to one-loop diagram contribution regardless of imposition of discrete or U(1) symmetry. In category (iii), the topological Lorentz structures T4-1-$ii$ and T4-2-$i$ are one-loop extensions of Type-II seesaw whereas T4-3-$i$ corresponds to one-loop extension of Type-I/III seesaw scenario.

\noindent The divergent topological diagrams (category (i)) require counter-terms to absorb the divergences which are in turn its tree-level realizations. The topological Lorentz structures of category (ii) such as T1-$i$ have been realized in Refs. \cite{Ma:2006km, Kubo:2006yx,Ma:2007kt}  while the well-studied Zee model \cite{Zee:1980ai} is a realization of T1-$ii$ Lorentz structure. Also, implications of Lorentz structures  T1-$iii$ and T3  have been discussed in Ref. \cite{Ma:1998dn}. In category (iii), the tree-level contribution to neutrino masses dominates. In Ref. \cite{Kanemura:2012rj}, the authors have discussed the possible realization of T4-1-$ii$ topological structure wherein the tree-level terms were inhibited using discrete symmetry and the neutrino masses were generated by dimension, $d=7$ operator.  One of $U(1)_{B-L}$ model proposed in Ref. \cite{Wang:2015saa} reduces to topology T4-3-$i$ with one-loop dominant contribution to neutrino masses. In Ref. \cite{Bonnet:2012kz}, it was suggested that for category (iii) topological Lorentz structures implementation of $Z_2$ symmetry and assuming fermion running in the loop to be of Majorana nature leads to one-loop dominant contribution to neutrino masses, provided all couplings conserves the lepton number. In this way one can forbids the tree-level terms effectively leading to dominant one-loop seesaw contributions. However, neutrino mass generation, in general, leads to lepton number violation in realistic models. It is difficult to realize T4-2-$i$  with dominant one-loop contribution since scalar triplet couples to lepton doublets leading to tree-level dominant contribution. In the existing literature, one such attempt has been made employing D$_4$, cyclic symmetries with enlarged field content wherein lepton number is violated by right-handed Majorana neutrino mass couplings \cite{Loualidi:2020jlj}.  Here, we argue that one may not require additional fields (fields other than required by the topology) to suppress the tree-level contribution if we work within the paradigm of modular symmetry. As an example, we have constructed a possible realization of T4-2-$i$ topology based on A$_4$ modular symmetry.

\noindent In this work, we propose a realization of one-loop topology T4-2-$i$  which, essentially, requires two scalar doublets ($\phi$, $\rho$), one scalar triplet ($\Delta$) and fermionic field(s) $\psi$.  Keeping in view the advantage of modular symmetries over discrete symmetries that Yukawa couplings transform like other scalar or fermionic fields, we realize the topology using the A$_4$ modular symmetry in a minimal way. The field content of the model includes fields permitted by the topology only which is more economical and minimalist than the model proposed in Ref. \cite{Loualidi:2020jlj}.  The lepton number violation is manifested by assuming fermions running inside the loop as right-handed Majorana neutrinos. The tree-level Dirac neutrino mass term (emanating from the coupling of Higgs field with lepton doublet and right-handed Majorana neutrino) and tree-level contribution from scalar triplet is inhibited, successfully, by assigning suitable weights under the A$_4$ modular symmetry. Consequently, neutrino masses are generated by one-loop Type-II seesaw without the use of additional beyond standard model (BSM) field(s). The dark matter candidate(s) running in the loop are stabilized by assigning odd modular weights under A$_4$ modular group. We analyse the viability of the model under current neutrino oscillation data. Also, considering the possibility of fermionic dark matter, we have obtained the prediction on relic density of dark matter consistent with upper bound on branching ratio of lepton flavor violating (LFV) process $\mu \rightarrow e \gamma$. Furthermore, the implications of the model for direct detection of dark matter are obtained.  

\noindent The paper is organized as follows. Section 2 is devoted for the  possible realization of T4-2-$i$ topological Lorentz structure. In this section, we first reproduce some general features of T4-2-$i$ topology to make context of the problem and then propose possible scenario for the suppression of tree-level couplings based on A$_4$ modular symmetry.  Further, we have illustrated possible realization of the model in supersymmetric (SUSY) framework.The numerical analysis based on the discussions presented in Section 2, has been carried out in Section 3. In Section 4, we discuss the framework of calculation for relic density, spin independent cross-section for direct detection of dark matter and lepton flavor violation. Finally, we brief the conclusions in Section 5. The preliminaries about the modular symmetry and its connection to the permutation groups are discussed in Appendix A.

\section{ $\Gamma_3\simeq A_4$ based T4-2-$i$ Model}

\noindent In this section, firstly we discuss the general features of  T4-2-$i$ topology to make relevant context of the problem. Also, using symmetry arguments it is shown that in topology T4 tree-level terms are always invariant and dominating regardless of imposition of discrete or U(1) symmetry. Secondly, we propose a realization of this topology by assigning suitable charge assignments to the field content and Yukawa couplings under A$_4$ modular invariance (see Appendix A for details).

\subsection{General features of T4-2-$i$ Topology} \label{3.1}
\begin{figure}
             \centering \includegraphics[height=2 in]{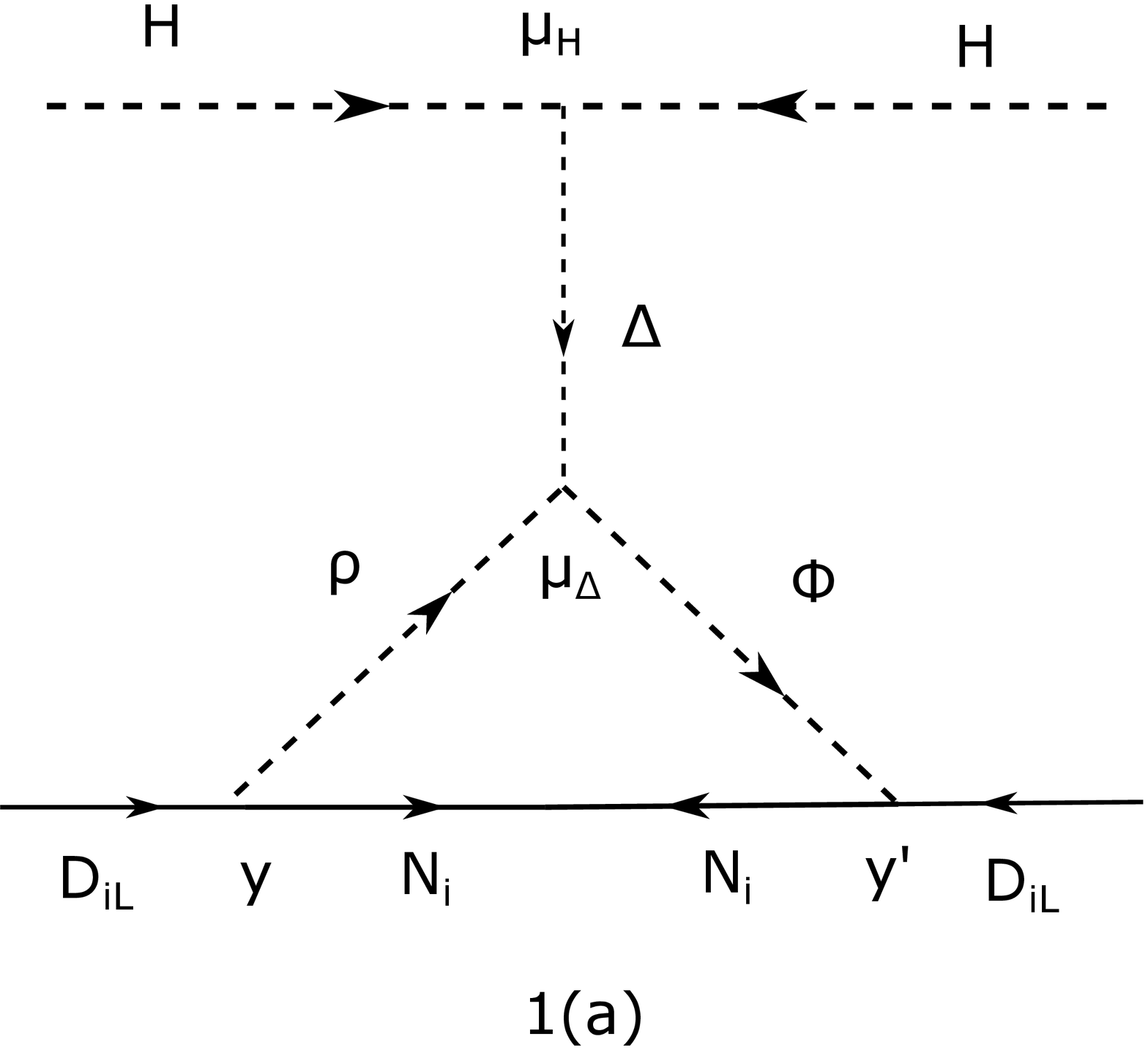} \includegraphics[height=2 in]{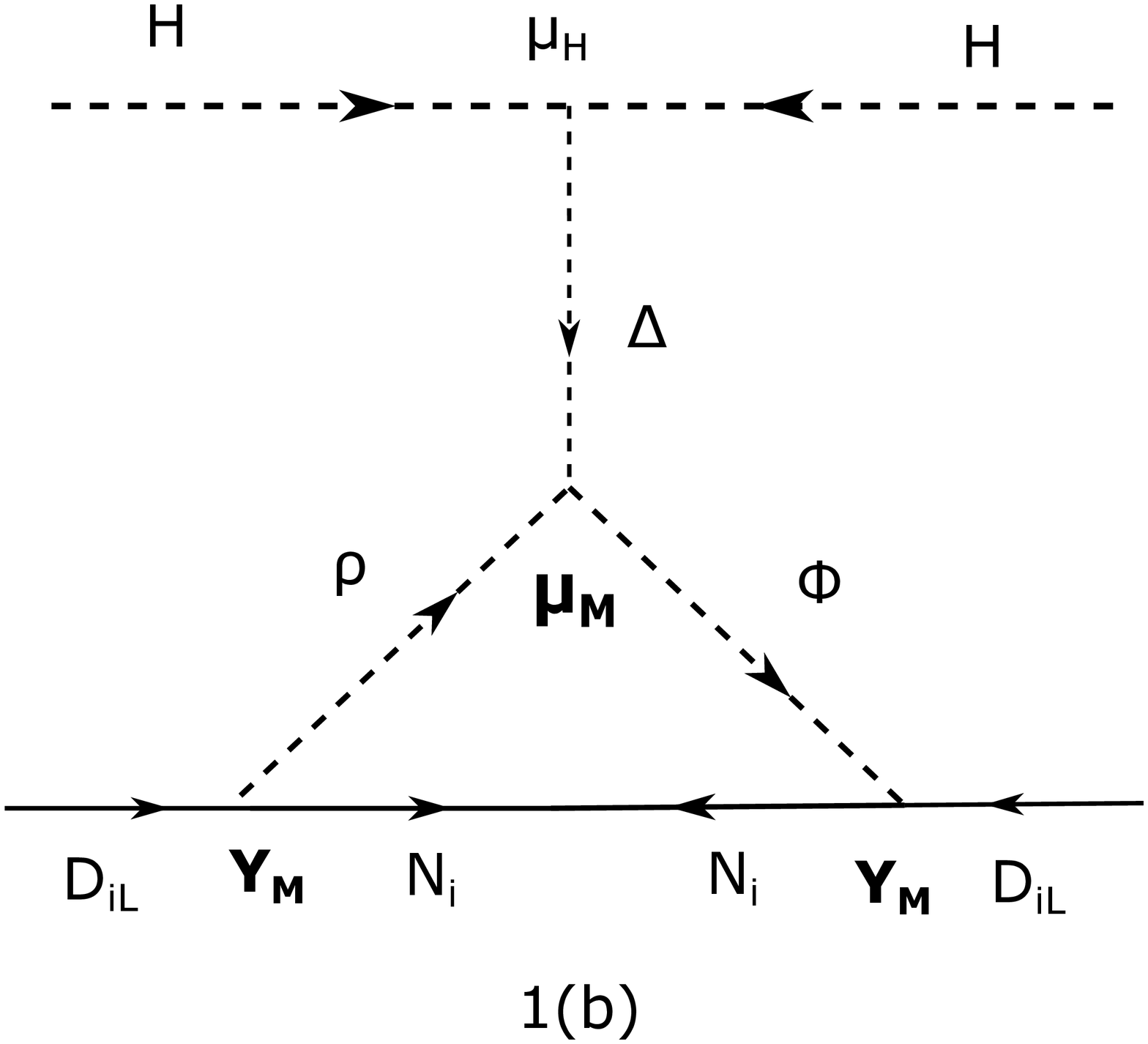} 
             \caption{General T4-2-i topological structure (Fig. 1(a)) wherein Yukawa couplings are constants and effect of A$_4$ modular symmetry to inhibit the tree-level contribution (Fig. 1(b)). The couplings $\bf{\mu_{H_M}}$, $\bf{\mu_{\Delta_M}}$, $\bf{Y_M}$ and $\bf{Y'_M}$ in Fig. 1(b) transform under the A$_4$ modular symmetry.}
             \label{fig:1}
         \end{figure}

In general, finite diagrams of T4 topology are one-loop extensions of canonical Type-I/II/III seesaw scenarios. Particularly, T4-2-$i$ topological Lorentz structure is one-loop extension of Type-II seesaw.  It extends the SM with two scalar doublets ($\rho,\phi$) and heavy Majorana right-handed neutrinos ($N_{i}$) as shown in Fig. \ref{fig:1}(a). It is to be noted that
\begin{enumerate}
\item If scalar doublet acquires the $vev$, one-loop diagram reduces to tree-level diagram of canonical Type-II seesaw.
    \item The tree-level contribution act as leading contribution in addition to one-loop contribution to neutrino masses.
\end{enumerate}

\noindent It is straightforward to understand it from Fig. \ref{fig:1}(a):
In order to have topology T4-2-$i$ at one-loop level, all the vertices should be allowed by invariance under the new symmetry. Let $Q_j$ be the quantum number under the new symmetry. For the vertex $\mu_{H} H \Delta^{\dagger} H$ to be allowed, we require $2Q_{H}+Q_{\Delta}=0$. Also, invariance of vertices $\mu_{\Delta}\Delta \rho  \phi^{\dagger}$, Yukawa couplings $y$ and $y'$  require  $$Q_{\Delta}-Q_{\rho}+Q_{\phi}=0,$$
$$Q_{D_{L}}+Q_{N_{i}}-Q_{\rho}=0,$$
$$Q_{D_{L}}-Q_{N_{i}}+Q_{\phi}=0,$$
respectively,  which, further reduces to $2Q_{D_{L}}+Q_{\Delta}=0$ implying that tree-level Type-II seesaw term is always present. It is evident from above analysis that if vertex $\mu_{H} H \Delta^{\dagger} H$ is allowed, tree-level term is also allowed. However, in order to have dominant one-loop contribution the tree-level term must be inhibited while the vertex $\mu_{H} H \Delta^{\dagger} H$ should be allowed under the new symmetry invariance.  Similar problem arises in other finite diagrams of T4 topological Lorentz structures.

\subsection{Implications of A$_4$ Modular Symmetry}\label{3.2}
 \noindent A$_4$ Modular symmetry have even modular weights for Yukawa couplings which are treated as constant in conventional flavor symmetric models. The matter fields can have integral modular weights such that sum of modular weights vanishes for an invariant term of Lagrangian (as discussed in Appendix A). \\
 \noindent In order to realize the topology T4-2-$i$ with dominant one-loop contribution, we need to ensure following:
\begin{enumerate}
\item There is no tree-level contribution from scalar triplet i.e. $2Q_{L}+Q_{\Delta}\neq 0$ while the vertex $\mu_{H} H \Delta^{\dagger} H$ should be allowed.
\item The Dirac neutrino mass term emanating from the coupling of Higgs field with lepton doublet and right-handed Majorana neutrino is inhibited. 
\item Essentially, the vertices with couplings $\mu_{\Delta}$, $y$ and $y'$(Fig. \ref{fig:1}(a) should be allowed such that  neutrino mass generation is manifested at one-loop level via scalar triplet only.
\end{enumerate}
\noindent Under A$_4$ modular paradigm, we employ fields permitted by the topology only i.e.  two scalar doublets($\phi$, $\rho$), one scalar triplet ($\Delta$) and assuming the fermions inside the loop as right-handed Majorana neutrinos. The field content of the model and corresponding modular weights are given in Table \ref{tab1}. Also, the transformations of Yukawa couplings under A$_4$ modular symmetry are shown in Table \ref{tab2}.\\

\begin{table}[t]
		\centering
		\begin{tabular}{cccccccccccccc}
			Symmetry & $\bar{D}_{eL}$  & $\bar{D}_{\mu L}$ & $\bar{D}_{\tau L}$& $e_{R}$ & $\mu_{R}$ & $\tau_{R}$ & $N_{1}$ & $N_2$&$N_3$ &$H$ &  $\Delta$  & $\tilde{\phi}$&$\rho$ \\ \hline
			$SU(2)_L$    &     2 &     2 &     2      &    1      &   1     &   1& 1       & 1 & 1 &    2     & 3 &2&2  \\ \hline
			$U(1)_Y$    &   $ 1    $ &   $ 1    $  &   $ 1   $  &   -2      &   -2     &   -2       & 0 & 0&0 &   1     &    2  & -1  & -1  \\  \hline
			$A_{4}$ &      1 &     1$''$ &     1$'$    &      1    &     1$'$      &     1$''$       & 1 & 1$'$ &1$''$ & 1 &    1   & 1    & 1  \\ \hline
			-$k_{I}$ & 2 & 2 & 2 &-2 & -2 & -2 & -3& -3&-3 & 0 &  0& -3 & -3 \\ 
		\end{tabular}
		\caption{\label{tab1} The field content of the model and respective charge assignments under $SU(2)_L$, $U(1)_Y$, A$_{4}$. including modular weights.} 
	\end{table}

\begin{table}
\begin{center}
\begin{tabular}{ccccccccc}
 $Y^{n}_{m}$&  $Y^{4}_{1}$  & $Y^{4}_{1'}$  &  $Y^{6}_{1}$ &  $Y^{8}_{1}$ &  $Y^{8}_{1'}$  &  $Y^{8}_{1''}$&  $Y^{10}_{1}$ &  $Y^{10}_{1'}$ \\
 \hline
 A$_4$&1& 1$'$ & 1  & 1&1$'$  &  1$''$ &1 & 1$'$  \\
 \hline
	-$k_{I}$ & 4 & 4 & 6 & 8 & 8 & 8 & 10 & 10
\end{tabular}

\caption{\label{tab2} The transformations of higher order Yukawa couplings under A$_{4}$ modular symmetry.}
\end{center}
\end{table}
\noindent {\textbf{\underline{Scalar potential terms:}}} Firstly, we assume Higgs field ($H$) and scalar triplet ($\Delta$) as trivial $A_4$ singlets with zero weight so that vertex $\mu_{H_M} H \Delta^{\dagger} H$ is allowed under $A_4$ modular symmetry. Secondly, the fields running inside the loop i.e. $N_i,\rho$ and $\phi$ are assigned singlet representation under A$_4$. Further, they have odd modular weights to ensure stability making them suitable dark matter candidates in the model (Table \ref{tab1})). The above assignments requires $\mu_{\Delta}$ mass coupling to have modular weight $6$ so that the vertex $\mu_{\Delta_M}\Delta \phi^{\dagger}\rho$ is allowed.\\
 
\noindent {\textbf{\underline{Inhibition of tree-level terms:}}}
 With the above charge  assignments, the right-handed Majorana neutrino mass terms are allowed if Yukawa couplings transform as singlet of modular weight $6$. Also, the charged lepton fields transform under A$_4$ modular symmetry such that charged lepton mass matrix is diagonal with modular weight $6$. Consequent to the above discussion:
 \begin{enumerate}
    \item The Yukawa couplings $y$ and $y'$ have modular weight -$4$ which can be made invariant if $Y_{M}$ and $Y'_{M}$ acquire modular weight +$4$ i.e. $Y^{4}_{1}$, $Y^{4}_{1'}$, as shown in Fig. \ref{fig:1}(b).
     \item The tree-level Dirac term is not allowed since sum of modular weights is odd.
     \item Also, the tree-level Type-II seesaw term is disallowed as sum of modular weight is positive and even.
 \end{enumerate}
 Hence, tree-level contribution is suppressed due to the imposition of A$_4$ modular invariance, thus, resulting in dominant one-loop contribution to neutrino mass.\\
 
 \noindent {\textbf{\underline{One-loop contribution to neutrino mass:}}} The Yukawa Lagrangian for charged leptons under A$_4$ modular invariance is given by
\begin{equation} \label{charged}
\mathcal{L}_{I}= \alpha_{l}(\bar{D}_{eL} H e_{R} ) +\beta_{l}(\bar{D}_{\mu L} H \mu_{R} )+\gamma_{l}(\bar{D}_{\tau L} H \tau_{R} ),
\end{equation}
where $\alpha_{l}$, $\beta_{l}$ and $\gamma_{l}$ are coupling constants. After the spontaneous symmetry breaking, the Higgs field acquires the vacuum expectation value (\textit{vev}), $v_{H}$. The resulting charged lepton mass matrix, $M_l$, is given by \\
\begin{equation} \label{charged_M}
	M_{l}=\frac{v_H}{\sqrt{2}}
	{\begin{pmatrix}
		\alpha_l& 0&0 \\
		0  & \beta_l& 0 \\
		0& 0 & \gamma_l  
		\end{pmatrix}}.
	\end{equation}
	With modular transformations given in Table \ref{tab1} and \ref{tab2}, the Yukawa Lagrangian describing the neutrino mass generation at one-loop is given by
	\begin{eqnarray} \label{neutrino}
\nonumber
\mathcal{L}_{II}&&= g_{1}(\bar{D}_{eL}  \rho N_{1} Y^{4}_{1}) + g_{2}(\bar{D}_{\mu L}  \rho N_{2} Y^{4}_{1})+ g_{3}(\bar{D}_{\tau L}  \rho N_{3} Y^{4}_{1})\\ \nonumber
&&+k_{1}(\bar{D}_{e L}  \rho   N_{3} Y^{4}_{1'})+k_{2}(\bar{D}_{\mu L} \rho N_{1} Y^{4}_{1'})+ k_{3}(\bar{D}_{\tau L}  \rho  N_{2} Y^{4}_{1'})\\
\nonumber&&+a_{1}(\bar{D}_{eL}  \tilde{\phi} N_{1} Y^{4}_{1}) + a_{2}(\bar{D}_{\mu L}  \tilde{\phi} N_{2} Y^{4}_{1})+ a_{3}(\bar{D}_{\tau L} \tilde{\phi} N_{3} Y^{4}_{1})\\ \nonumber
&&+b_{1}(\bar{D}_{e L}  \tilde{\phi}   N_{3} Y^{4}_{1'})+b_{2}(\bar{D}_{\mu L} \tilde{\phi}N_{1} Y^{4}_{1'})+ b_{3}(\bar{D}_{\tau L}   \tilde{\phi}  N_{2} Y^{4}_{1'})\\
&&+ M'_{1} \bar{N^{c}_{1}} N_{1} Y^{6}_{1} + M'_{2} (\bar{N^{c}_{2}} N_{3} Y^{6}_{1}+\bar{N^{c}_{3}} N_{2} Y^{6}_{1}),
\end{eqnarray}
where $g_{i},k_{i},a_{i}$ and $b_{i}$ ($i=1,2,3$) are coupling constants.
Here, modular weights play the role of $Z_2$ symmetry stabilizing the dark matter candidates and inhibiting tree-level Yukawa couplings. The odd modular weights act as $Z_2$ odd charges which stabilizes the fields dictated by topology itself i.e. scalar doublets ($\phi,\rho$) and right-handed Majorana neutrinos ($N_{1},N_2,N_3$)  running in the loop. The scalar doublets ($\phi,\rho$) are inert having no vacuum expectation value ($vev$). The Lagrangian in Eqn. (\ref{neutrino}) results in the following Dirac Yukawa matrices,
\begin{equation} \label{dirac}
	y_{\rho}=
	{\begin{pmatrix}
	g_{1} Y^{4}_{1} &0& k_{1} Y^{4}_{1'}\\
    k_{2} Y^{4}_{1'}& g_{2} Y^{4}_{1}  &0 \\
    0 & k_{3} Y^{4}_{1'} &g_{3} Y^{4}_{1}
		\end{pmatrix}},\hspace{0.2cm}
			y_{\phi}=
	{\begin{pmatrix}
	a_{1} Y^{4}_{1} &0& b_{1} Y^{4}_{1'}\\
	b_{2} Y^{4}_{1'}&a_{2} Y^{4}_{1} & 0 \\
    0& b_{3} Y^{4}_{1'}&	a_{3} Y^{4}_{1}  
		\end{pmatrix}}.
	\end{equation}
	Also, the right-handed Majorana neutrino mass matrix $M_R$ is given by
\begin{equation} \label{right}
	M_{R}=
	{\begin{pmatrix}
	M'_{1} Y^{6}_{1} & 0&0\\
		0 & 0&M'_{2} Y^{6}_{1}\\
		0&M'_{2} Y^{6}_{1}&0
		\end{pmatrix}},
		\end{equation}
where $M'_{k}$ $(k=1,2)$ are right-handed neutrino mass scales of the bare mass terms. The right-handed neutrino masses $M_{k}$ $(k=1,2,3)$ can be obtained by diagonalizing $M_{R}$ using the unitary mixing matrix $U_{R}$ i.e. \textit{diag}$(M_{1}, M_{2},
M_{3})=U_{R}M_{R}U_{R}^{T}$. In $M_{R}$-diagonal basis, the Dirac Yukawa matrices (Eqn. (\ref{dirac})) are given by  
\begin{equation}\label{modifieddirac}
    Y_{\rho}=y_{\rho} U_{R}; \hspace{1cm} Y_{\phi}=y_{\phi} U_{R}.
\end{equation}
Using the Eqns. (\ref{dirac}-\ref{modifieddirac}), the dominant one-loop contribution to neutrino mass is given by \cite{Bonnet:2012kz}
\begin{equation}\label{mnuformula}
    M_{\nu}=-\sum_{k}\frac{\mu_{\Delta_M}v^{2}}{m_{\Delta}^2}\mu_{H_M}[Y_{\rho} (M_{k}) Y_{\phi}^{T}] \mathcal{I}(M_{\rho}^{2},M_{\phi}^{2},M_{k}^2),
\end{equation}
    where $\mu_{\Delta_M}$ and $\mu_{H_M}$ are couplings at vertices $\mu_{H_M} H \Delta^{\dagger} H$ and $\mu_{\Delta_M}\Delta   \phi^{\dagger}\rho$(Fig. \ref{fig:1}(b)), respectively. The loop factor in the Eqn. (\ref{mnuformula}) is given by
    \begin{eqnarray} \label{loopfactor}
\nonumber
 \mathcal{I}(M_{\rho}^{2},M_{\phi}^{2},M_{k}^2)=&&-\Bigg(\frac{1}{4\pi} \frac{M_{\rho}^{2}}{(M_{\rho}^{2}-M_{\phi}^{2})(M_{\rho}^{2}-M_{k}^{2})} \ln \frac{M_{k}^{2}}{M_{\rho}^{2}}\\ 
&&+\frac{M_{\phi}^{2}}{(M_{\phi}^{2}-M_{\rho}^{2})(M_{\phi}^{2}-M_{k}^{2})} \ln \frac{M_{k}^{2}}{M_{\phi}^{2}}\Bigg).
\end{eqnarray}
\noindent The neutrino mass matrix emanating from Eqn. (\ref{mnuformula}) is, in general complex and asymmetric. In order to have complex symmetric $M_\nu$, the coupling constants must satisfy $g_{i}=a_{i}$ and $k_{i}=b_{i}$ ($i=1,2,3$) equalities. It is to noted that we have not considered the non-trivial relations amongst Yukawa couplings (as discussed in Ref. \cite{Loualidi:2020jlj}) because, here, they transform under the A$_4$ modular symmetry. Also, it is to be noted that, the modular weight of loop particles \textit{viz.} right-handed Majorana neutrinos ($N_{i}$), inert doublets ($\rho, \phi$) is odd and thus, are suitable candidate for dark matter. Here, we have considered one such possibility, fermionic dark matter, by assuming mass of right-handed neutrino ($M_1$)  to be smallest. \\
\noindent Thus, the neutrino mass matrix, $M_{\nu}$, is proportional to $Y_{\rho} Y_{\phi}^{T}$ such that
\begin{equation}\label{scalemnu}
    M_{\nu}=\mathcal{K}(Y_{\rho} Y_{\phi}^{T})\equiv \mathcal{K} \tilde{M_{\nu}},
\end{equation}
where $\mathcal{K}=M_{1} \mathcal{I}(M_{\rho}^{2},M_{\phi}^{2},M_{1}^2)$ is the overall scale factor. The neutrino masses are given by $m_{i}=\mathcal{K}\tilde{m_{i}}(i=1,2,3)$, where $\tilde{m_{i}}$ are mass eigenvalues of $Y_{\rho} Y_{\phi}^{T}$. For Normal hierarchy, the scale factor is obtained using the atmospheric mass-squared difference $\Delta m_{31}^2$ (Table \ref{tab5}) as 
\begin{equation}\label{dms31}
   \mathcal{K}^2=\frac{\Delta m_{31}^2}{(\tilde{m_{3}}^{2}-\tilde{m_{1}}^{2})}.
\end{equation}
Consequently, the solar mass-squared difference, in terms of $\mathcal{K}$, can be written as
\begin{equation}\label{dms21}
\Delta m_{21}^2 = \mathcal{K}^2 (\tilde{m_{2}}^{2}-\tilde{m_{1}}^{2}).
\end{equation}
The neutrino mixing matrix $U$ is obtained by diagonalizing $M_{\nu}$ using the transformation $U^{T}M_{\nu}U$=\textit{diag}$(m_{1},m_{2},m_{3})$. Since charged lepton mass matrix $M_{l}$ (Eqn. (\ref{charged_M})) is diagonal, the lepton mixing matrix is equal to neutrino mixing matrix  i.e. $U_{PMNS}=U$, where $U_{PMNS}$ is Pontecorvo–Maki–Nakagawa–Sakata matrix. The  mixing angles can be evaluated using elements of the neutrino mixing matrix, $U$, as
 \begin{eqnarray} \label{mixingangles}
 &&\quad \sin ^{2} \theta_{13}=\left|U_{13}\right|^{2}, \quad \sin ^{2} \theta_{12}=\frac{\left|U_{12}\right|^{2}}{1-\left|U_{13}\right|^{2}}, \quad \sin ^{2} \theta_{23}=\frac{\left|U_{23}\right|^{2}}{1-\left|U_{13}\right|^{2}}.
\end{eqnarray}

\noindent The Jarlskog rephasing $CP$ invariant\cite{Jarlskog:1985ht,Krastev:1988yu} is given by  
\begin{equation} \label{jcp}
J_{C P}=\operatorname{Im}\left[U_{11} U_{22} U^{*}_{1 2} U^{*}_{2 1}\right],
\end{equation}

\noindent while other two $CP$ invariants \( I_{1} \) and \( I_{2} \) related to Majorana phases ($\alpha$, $\beta$) are

\begin{equation}\label{i1i2}
I_{1}=\operatorname{Im}\left[U_{11}^{*} U_{12}\right], \hspace{1 cm} I_{2}=\operatorname{Im}\left[U_{11}^{*} U_{13}\right].
\end{equation}
Another important parameter to investigate is the effective Majorana neutrino mass($M_{ee}$) which can shed light on the nature of neutrino being Dirac or Majorana particle. The effective Majorana mass is given by
\begin{equation}\label{Mee}
 M_{e e} =\left| \sum_{i=1}^{3}U_{1i}m_{i}\right|.
\end{equation}
The viability of model with neutrino oscillation data(Table \ref{tab5}) and its predictions for $M_{ee}$ and $CP$ violation is discussed in the next section.
\subsection{Supersymmetric (SUSY) realization of the Model}
\begin{table}[t]
		\centering
		\begin{tabular}{cccccccccc}
			Symmetry & $\hat{D}_{eL}$  & $\hat{D}_{\mu L}$ & $\hat{D}_{\tau L}$& $e_{R}^{c}$ & $\mu_{R}^{c}$ & $\tau_{R}$ & $N_{1}^{c}$ & $N_2^{c}$&$N_3^{c}$ \\ \hline
			$SU(2)_L$    &     2 &     2 &     2      &    1      &   1     &   1& 1       & 1 & 1 \\ \hline
			$U(1)_Y$    &   $ -1    $ &   $ -1    $  &   $ -1   $  &   -2      &   -2     &   -2       & 0 & 0&0   \\  \hline
			$A_{4}$ &      1 &     1$''$ &     1$'$    &      1    &     1$'$      &     1$''$       & 1 & 1$'$ &1$''$  \\ \hline
			-$k_{I}$ & 2 & 2 & 2 &-2 & -2 & -2 & -3& -3&-3  \\ 
		\end{tabular}
		\caption{\label{tab3} The superfield content of the model and respective charge assignments under $SU(2)_L$, $U(1)_Y$, A$_{4}$ including modular weights.} 
	\end{table}

 \begin{table}[t]
		\centering
		\begin{tabular}{ccccccccc}
			Symmetry & $H_u$ &  $H_d$  & $\phi_u$ & $\phi_d$ & $\rho_u$ & $\rho_d$ & $\Delta_{u}$ & $\Delta_{d}$ \\ \hline
			$SU(2)_L$    &     2 &     2 &     2      &    2      &   2     &   2   & 3 &3  \\ \hline
			$U(1)_Y$    &  1 &  -1   &    1    &   -1      &   1     &   -1       & -2 & 2   \\  \hline
			$A_{4}$ &      1 &     1&     1   &      1    &     1      &     1       & 1 & 1 \\ \hline
			-$k_{I}$ & 0 & 0 & -3 &-3 & -3 & -3 & 0& 0\\ 
		\end{tabular}
		\caption{\label{tab4} The superfield content of the scalar sector of the model and respective charge assignments under $SU(2)_L$, $U(1)_Y$, A$_{4}$ including modular weights.} 
	\end{table}

The superfield in the $\theta$-expansion is given by \cite{Wess:1992cp}
\begin{equation}
    \Phi=\phi_{m}+\sqrt{2}\theta.\psi_{m}+\theta^{2}F_{m},
\end{equation}
where, $\phi_m$ is the scalar field, $\theta$ is the Grassmann variable, $\psi_m$ is spinor field and $F_m$ is auxiliary field. Since, SUSY transformation commutes with gauge transformation, superfields have same quantum numbers under the SM gauge group as shown in Table \ref{tab3}. The superfields $\hat{D}_{iL}$ are defined as left chiral superfield  lepton doublet, $i_{R}^{c}$ are right-handed $CP$-conjugated charged lepton superfields where $i=e, \mu, \tau$. The $CP$-conjugated right-handed neutrino fields are defined as $N_{j}^{c}$ where $j=1,2,3$. The SUSY invariant interactions of superfields are obtained from the holomorphic terms (F-terms) of the superpotential which do not contain the $H^{\dagger}$ terms leading to massless fermions. The second Higgs field $H_d=(H_{d}^{0},H_{d}^{-})$ play the role of $H^{\dagger}$ while $H_{u}=(H_{u}^{+},H_{u}^{0})$ is the usual SM Higgs field. Here, $d,u$ represent corresponding field giving mass to down-type and up-type quarks. Therefore, scalar sector has twice the number of fields in SM as shown in Table \ref{tab4}. The superpotential responsible for the masses of charged lepton is given by
\begin{equation} \label{charged1}
\mathcal{W}_{I}= \alpha_{l}(\hat{D}_{eL} H_{d} e_{R}^{c} ) +\beta_{l}(\hat{D}_{\mu L} H_{d} \mu_{R}^{c} )+\gamma_{l}(\hat{D}_{\tau L} H_{d} \tau_{R}^{c} ),
\end{equation}
where $\alpha_{l}$, $\beta_{l}$ and $\gamma_{l}$ are coupling constants. Also, the superpotential responsible for neutrino mass generation is given by 
\begin{eqnarray} \label{neutrino1}
\nonumber
\mathcal{W}_{II}&&= g_{1}(\hat{D}_{eL}  \rho_{d} N_{1}^{c} Y^{4}_{1}) + g_{2}(\hat{D}_{\mu L}  \rho_{d} N_{2}^{c} Y^{4}_{1})+ g_{3}(\hat{D}_{\tau L}  \rho_{d} N_{3}^{c} Y^{4}_{1})\\ \nonumber
&&+k_{1}(\hat{D}_{e L}  \rho_{d}   N_{3}^{c} Y^{4}_{1'})+k_{2}(\hat{D}_{\mu L} \rho_{d} N_{1}^{c} Y^{4}_{1'})+ k_{3}(\hat{D}_{\tau L}  \rho_{d}  N_{2}^{c} Y^{4}_{1'})\\
\nonumber&&+a_{1}(\hat{D}_{eL}  \phi_{d} N_{1}^{c} Y^{4}_{1}) + a_{2}(\hat{D}_{\mu L}  \phi_{d} N_{2}^{c} Y^{4}_{1})+ a_{3}(\hat{D}_{\tau L} \phi_{d} N_{3} Y^{4}_{1})\\ \nonumber
&&+b_{1}(\hat{D}_{e L}  \phi_{d}   N_{3}^{c} Y^{4}_{1'})+b_{2}(\hat{D}_{\mu L} \phi_{d} N_{1}^{c} Y^{4}_{1'})+ b_{3}(\hat{D}_{\tau L}   \phi_{d}  N_{2}^{c} Y^{4}_{1'})\\
&&+ M'_{1} N_{1}^{c} N_{1}^{c} Y^{6}_{1} + M'_{2} (N_{2}^{c} N_{3}^{c} Y^{6}_{1}+N_{3}^{c} N_{2}^{c} Y^{6}_{1}),
\end{eqnarray}
where $g_{i},k_{i},a_{i}$ and $b_{i}$ ($i=1,2,3$) are coupling constants. The SUSY breaking leads to diagonal charged lepton mass matrix (Eqn. (\ref{charged_M})) and Yukawa matrices as shown in Eqn. (\ref{dirac}). 

\noindent The vertex $\mu_{H_M} H_{d} \Delta_{d} H_{d}$ and $\mu_{\Delta_M}\Delta_{d}   \phi_{d}\rho_{d}$ are invariant under the $A_4$ modular symmetry with $\mu_{H_M}$ and $\mu_{\Delta_M}$  couplings transforming as $ Y^{6}_{1}$. Hence topology T4-2-$i$ is made functioned at one-loop level. The scalar potential of the model contains supersymmetric contribution from F-terms, D-terms and soft SUSY breaking terms, defined as 
\begin{equation}
    V=V_{SUSY}+V_{soft},
\end{equation}
where F term comprises of  $|F_{u}|^{2}$, $|F_{d}|^{2}$, $|F_{\Delta_{u}}|^{2}$, $|F_{\Delta_{d}}|^{2}$, $|F_{\rho_{u}}|^{2}$,$|F_{\rho_{d}}|^{2}$, $|F_{\phi_{u}}|^{2}$, $|F_{\phi_{d}}|^{2}$ and D-term contains $\vec{D}^{2}$ and $D^{2}$ which are given in Appendix B. The soft SUSY terms are given by

\begin{eqnarray}
    \nonumber
    V_{soft}=&&m_{H_u}^{2}|H_{u}|^{2}+m_{H_d}^{2}|H_{d}|^{2}+m_{\Delta_d}^{2}|\Delta_{d}|^{2}+m_{\Delta_u}^{2}|\Delta_{u}|^{2}+\mu_{\rho_{d}}|\rho_{d}|^2+\mu_{\rho_{u}}|\rho_{u}|^2+\mu_{\phi_{d}}|\phi_{d}|^2+\mu_{\phi_{u}}|\phi_{u}|^2 \\ \nonumber
    &&+(\mu_{1}^2 H_{u} H_{d}+\mu_{2}^2 \phi_{u} \phi_{d}+\mu_{3}^2 \rho_{u} \rho_{d}+\mu_{4}^2 \rho_{u} \phi_{d}+\mu_{5}^2 \phi_{u} \rho_{d} + h.c.)+\mu_{\Delta}Tr(\Delta_{u}\Delta_{d})\\ 
    &&+(\mu_{6} H_{u}\Delta_{u}H_{u}+\mu_{7} H_{d}\Delta_{d}H_{d}+\mu_{8} \rho_{u}\Delta_{u}\phi_{u}+\mu_{9} \rho_{d}\Delta_{d}\phi_{d}+ h.c.).
\end{eqnarray}
The soft SUSY breaking leads to the \textit{vev} of $H_{u,d}$, $\Delta_{u,d}$ giving charged lepton masses and neutrino masses at one-loop through inert scalar fields $\phi$ and $\rho$ having zero $vev$.
\section{Numerical Analysis}
\begin{table}[t]
\begin{center}
\begin{tabular}{c|c|c}
\hline \hline 
Parameter & Best fit $\pm$ \( 1 \sigma \) range & \( 3 \sigma \) range  \\
\hline \multicolumn{2}{c} { Normal neutrino mass ordering \( \left(m_{1}<m_{2}<m_{3}\right) \)} \\
\hline \( \sin ^{2} \theta_{12} \) & $0.304^{+0.013}_{-0.012}$ & \( 0.269-0.343 \)  \\
\( \sin ^{2} \theta_{13} \) & $0.02221^{+0.00068}_{-0.00062}$ & \( 0.02034-0.02420 \) \\
\( \sin ^{2} \theta_{23} \) & $0.570^{+0.018}_{-0.024}$ & \( 0.407-0.618 \)  \\
\( \Delta m_{21}^{2}\left[10^{-5} \mathrm{eV}^{2}\right] \) & $7.42^{+0.21}_{-0.20}$& \( 6.82-8.04 \) \\
\( \Delta m_{31}^{2}\left[10^{-3} \mathrm{eV}^{2}\right] \) & $+2.541^{+0.028}_{-0.027}$ & \( +2.431-+2.598 \) \\
\hline \multicolumn{2}{c} { Inverted neutrino mass ordering \( \left(m_{3}<m_{1}<m_{2}\right) \)} \\
\hline \( \sin ^{2} \theta_{12} \) & $0.304^{+0.013}_{-0.012}$ & \( 0.269-0.343 \)\\
\( \sin ^{2} \theta_{13} \) & $0.02240^{+0.00062}_{-0.00062}$ & \( 0.02053-0.02436 \) \\
\( \sin ^{2} \theta_{23} \) & $0.575^{+0.017}_{-0.021}$& \( 0.411-0.621 \) \\
\( \Delta m_{21}^{2}\left[10^{-5} \mathrm{eV}^{2}\right] \) & $7.42^{+0.21}_{-0.20}$ & \( 6.82-8.04 \) \\
\( \Delta m_{32}^{2}\left[10^{-3} \mathrm{eV}^{2}\right] \) & $-2.497^{+0.028}_{-0.028}$ & \( -2.583--2.412 \)  \\
\hline \hline
\end{tabular}
\end{center}
\caption{\label{tab5} Neutrino oscillation data from NuFIT 5.0 used in the numerical analysis \cite{Esteban:2020cvm}.}
\end{table}	
\begin{figure}[t]
	\begin{center}
 \includegraphics[height=6.0cm,width=7.5cm]{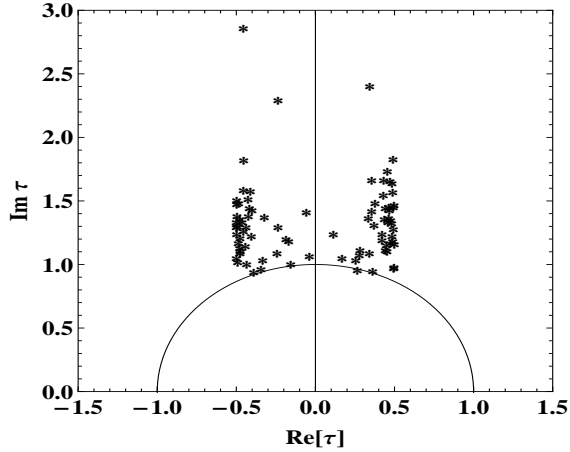}
		\end{center}
\caption{\label{fig:2} The parameter space of real and imaginary parts of complex modulus $\tau$ within the fundamental domain.}
\end{figure}
\noindent  In Eqn. (\ref{scalemnu}), apart from the common scale factor the neutrino mass matrix is function of $g_{i}$, $k_{i}$ ($i=1,2,3$) and Yukawa couplings having modular weight $4$. In general, the Yukawa couplings of higher modular weight ($4,6,8...$) are dependent on Yukawa couplings of modular weight $2$ which in turn is function of complex modulus $\tau$. In numerical analysis, the coupling constants, real and imaginary parts of $\tau$ and masses of BSM fields are varied randomly within the ranges given by

\begin{eqnarray}\label{paramters}
    \nonumber
    &&g_{i} \in [0.01,1];\hspace{0.2cm} k_{i} \in [0.01,1]; \hspace{0.2cm} |Re (\tau)| \in [0,0.5]; \hspace{0.2cm} Im (\tau) \in [0,1], \\ \nonumber
    &&M_{k} \in [1, 10^{3}] \text{ GeV } (M_{1}<M_{2}<M_{3});\hspace{0.2cm}M_{\rho,\phi} \in [1, 5] \text{ TeV},
\end{eqnarray}

\begin{figure}[]
	\begin{center}
			{\epsfig{file=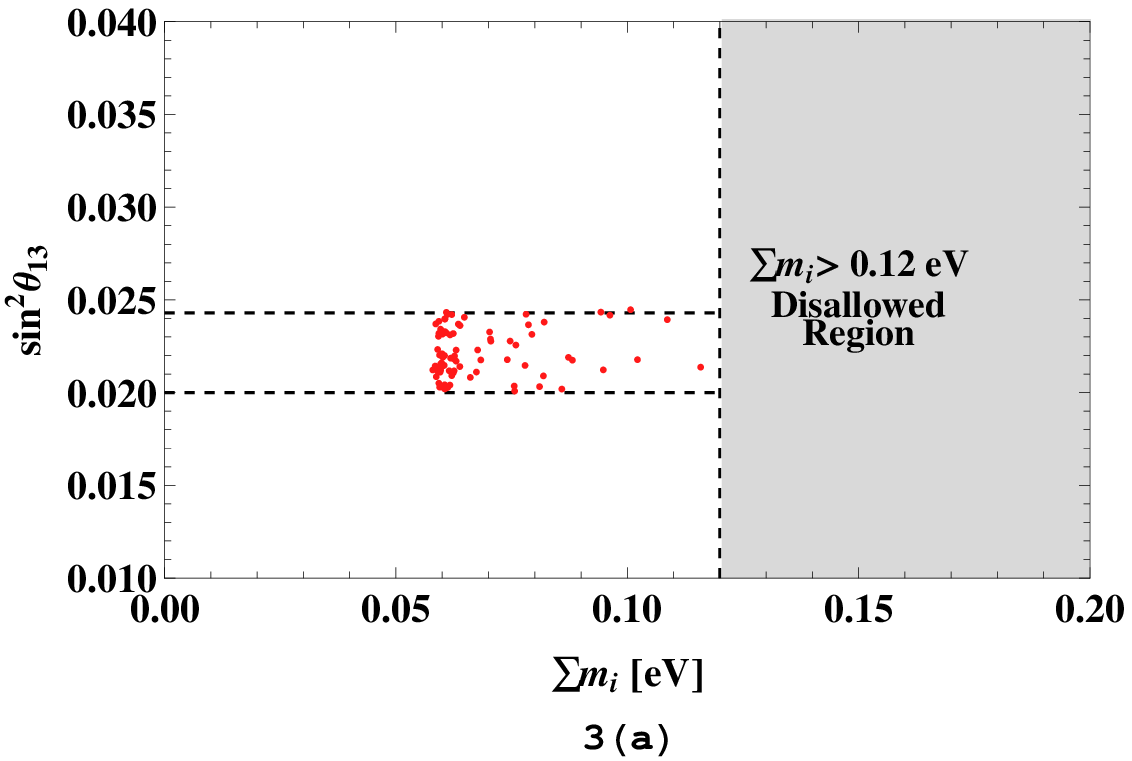,height=6.0cm,width=7.5cm}
				\epsfig{file=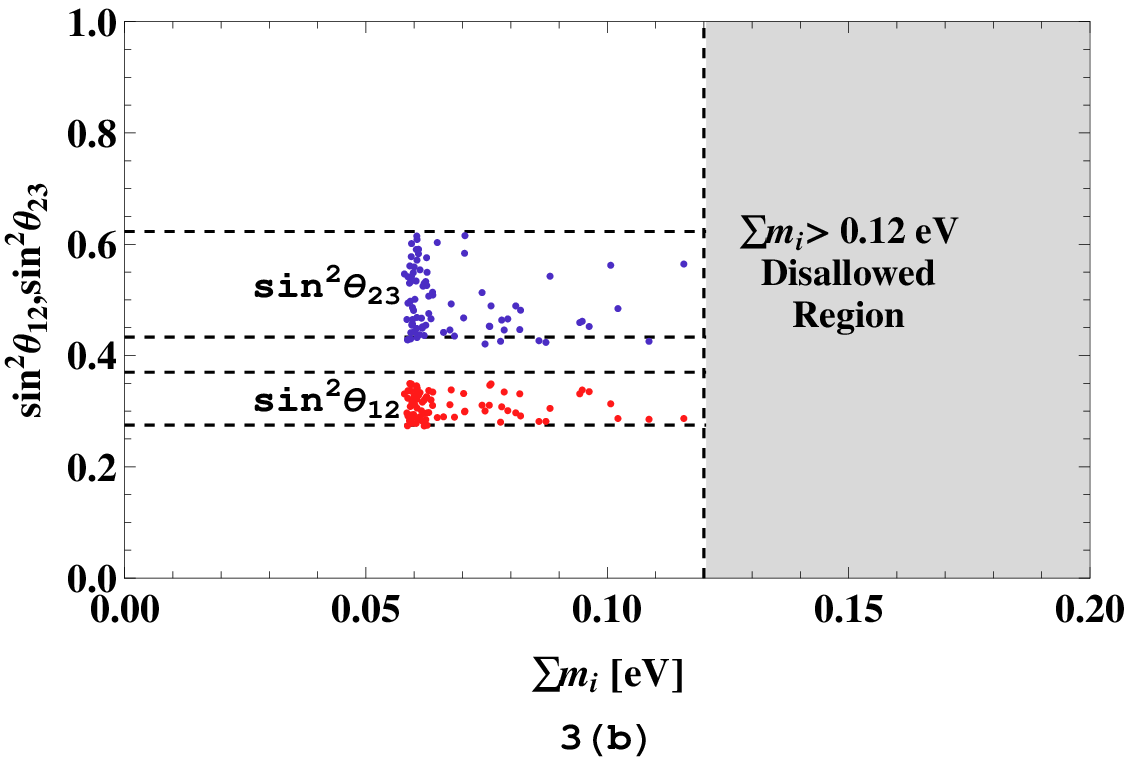,height=6.0cm,width=7.5cm}}
		\end{center}
\caption{\label{fig:3}  The variation of neutrino mixing angles with sum of neutrino masses $\sum m_{i}$  for Normal hierarchy. The grey shaded region is disallowed by cosmological bound on sum of neutrino masses\cite{Giusarma:2016phn,Aghanim:2018eyx}. The horizontal lines represent $3\sigma$ ranges of the mixing angle(Table \ref{tab5}).}
\end{figure} 
\begin{figure}[t]
	\begin{center}
			\epsfig{file=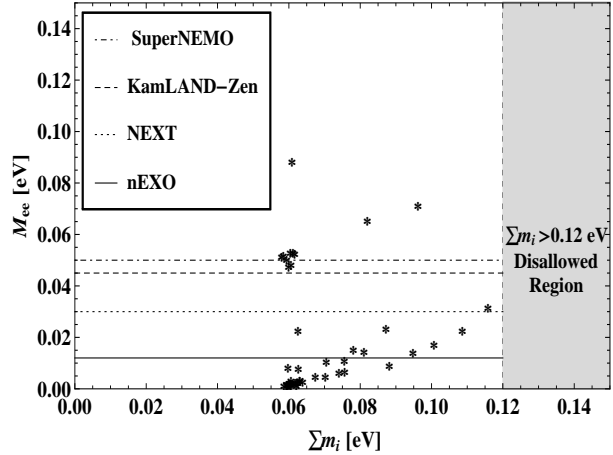,height=6.0cm,width=8.0cm}
		\end{center}
\caption{\label{fig:4} The variation of Effective Majorana mass $M_{ee}$ with  sum of neutrino masses $\sum m_{i}$  for normal hierarchical neutrino masses. The horizontal lines are the projective sensitivities of the $0\nu\beta\beta$ decay experiments. The grey shaded region is disallowed by cosmological bound on sum of the neutrino masses\cite{Giusarma:2016phn,Aghanim:2018eyx}.}
\end{figure}
\begin{figure}[h]
 \begin{center}
 \epsfig{file=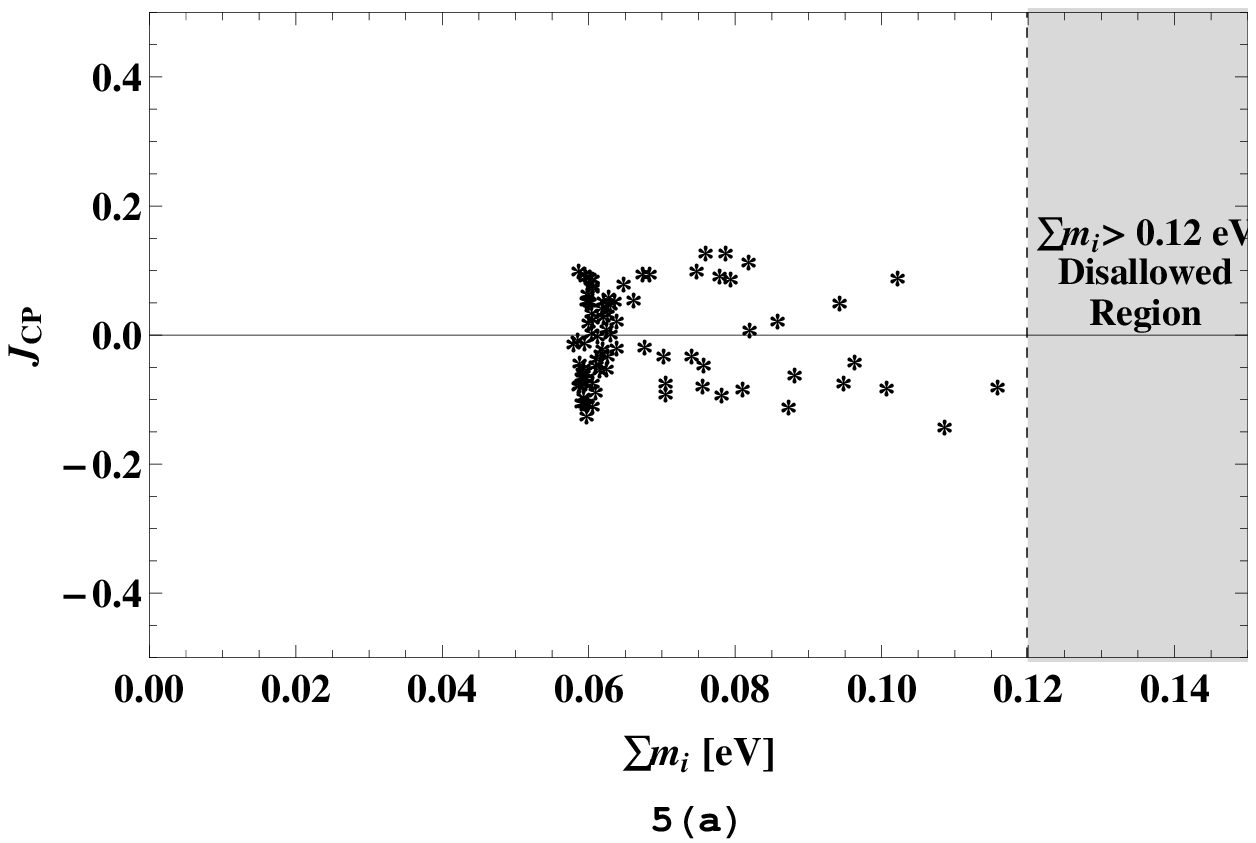,height=6.0cm,width=8.0cm}\\
 \vspace{0.5 cm}
{\epsfig{file=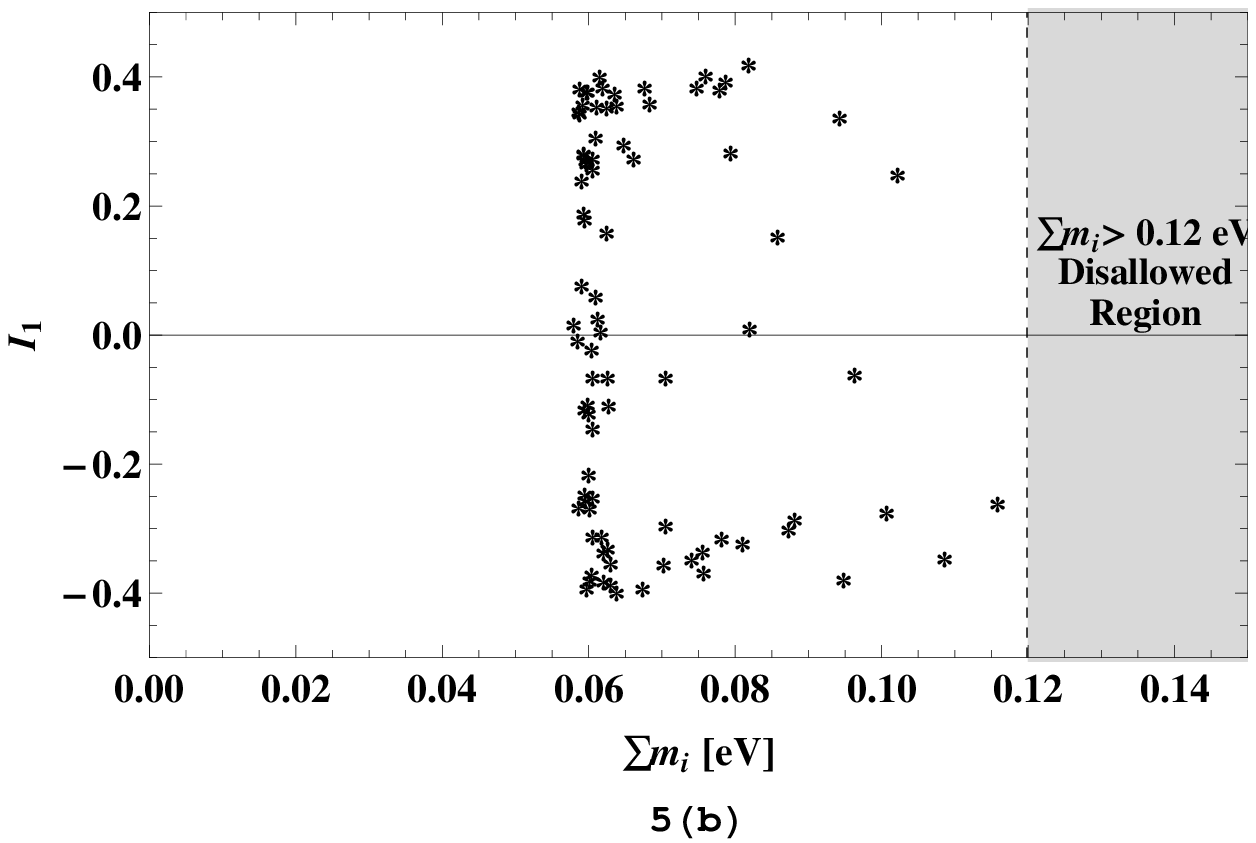,height=6.0cm,width=7.0cm},
\epsfig{file=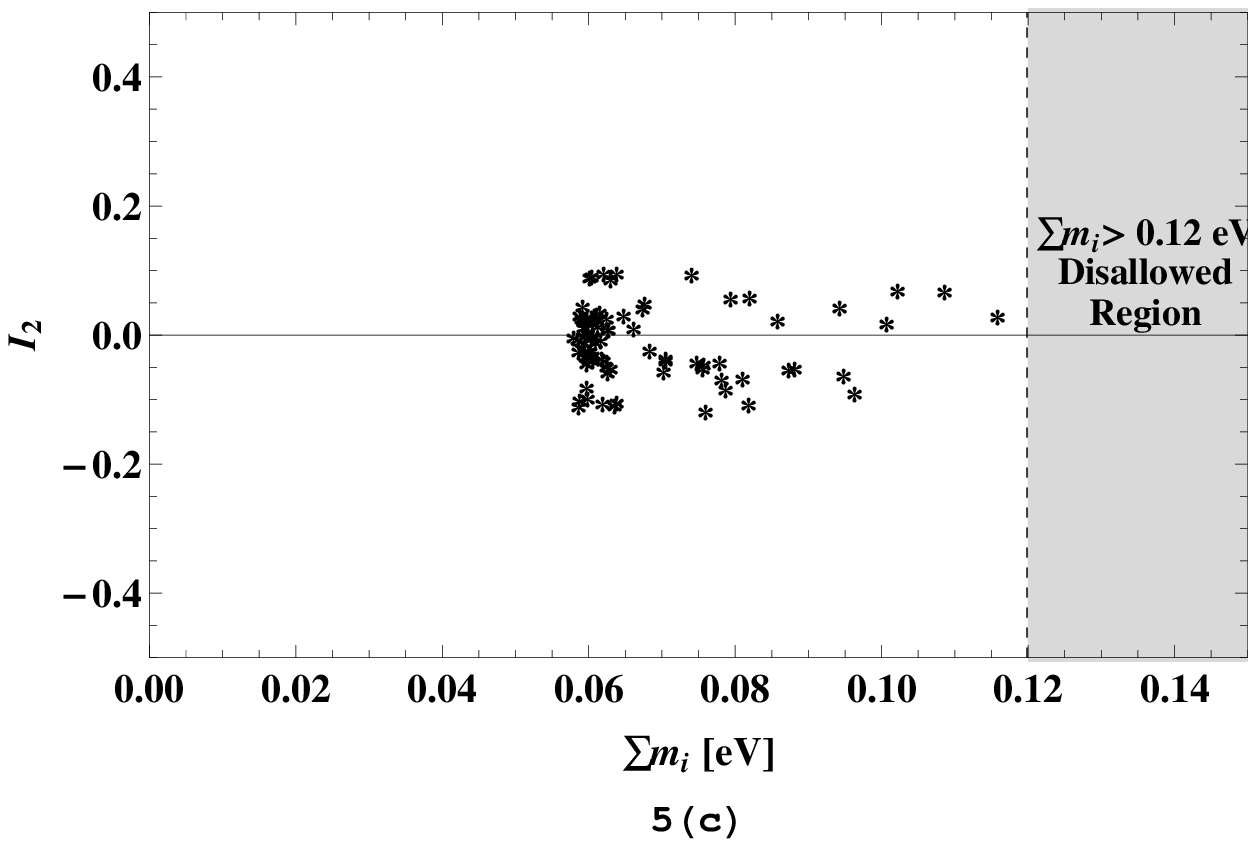,height=6.0cm,width=7.0cm}}
\end{center}
  \caption{\label{fig:5}The variation of  $CP$ invariants  $J_{CP}$,  $I_1$ and $I_2$  with sum of neutrino masses $\sum m_{i}$.} 
\end{figure}

\noindent to diagonalize $\tilde{M_{\nu}}$ (defined in Eqn. (\ref{scalemnu})), giving the mass eigenvalues $\tilde{m_{i}}$. The scale factor $\mathcal{K}$ is obtained on comparison with atmospheric mass-squared difference ($\Delta m_{31}^2$), using the Eqn. (\ref{dms31}). Using the scale factor, the neutrino mass-eigenvalues $m_{i}=\mathcal{K} \tilde{m_{i}}$  and corresponding neutrino mixing angles are evaluated using Eqn. (\ref{mixingangles}). We have used experimental constraints on neutrino mixing angles ($\theta_{12},\theta_{23},\theta_{13}$) and solar mass-squared difference ($\Delta m_{21}^2$) to ascertain the allowed parameter space of the model. In addition, the Yukawa couplings obey the perturbative constraint $Y \leq \sqrt{4\pi}$ \cite{Nomura:2019lnr}.

\begin{figure}[t]
\begin{center}	
\includegraphics[height=6.0cm,width=7.5cm]{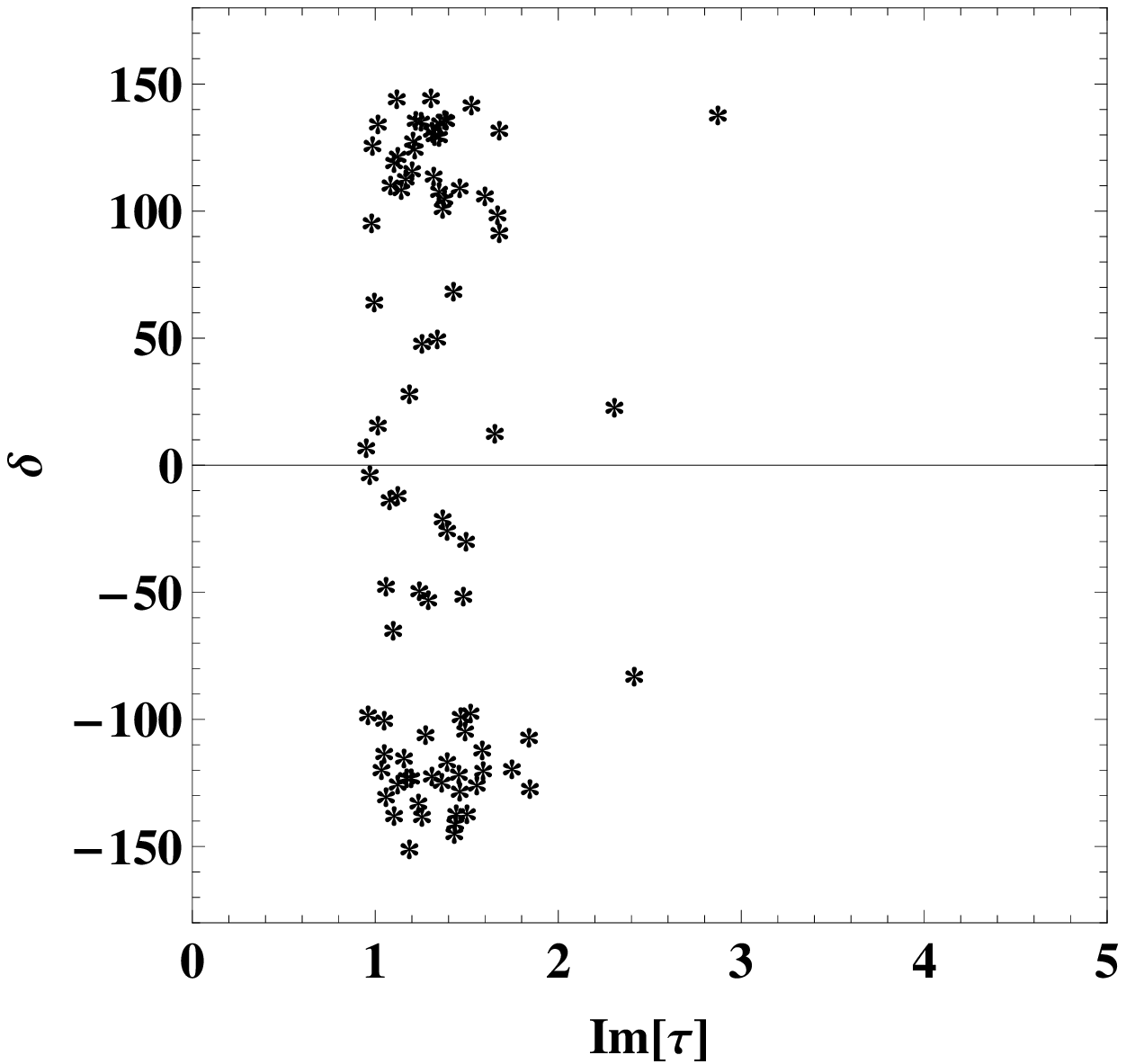}
\end{center}
    \caption{The variation of Dirac $CP$ phase $\delta$ with the imaginary part of complex modulus $\tau$.}
    \label{fig:6}
\end{figure}

\begin{figure}
	\begin{center}
			{\epsfig{file=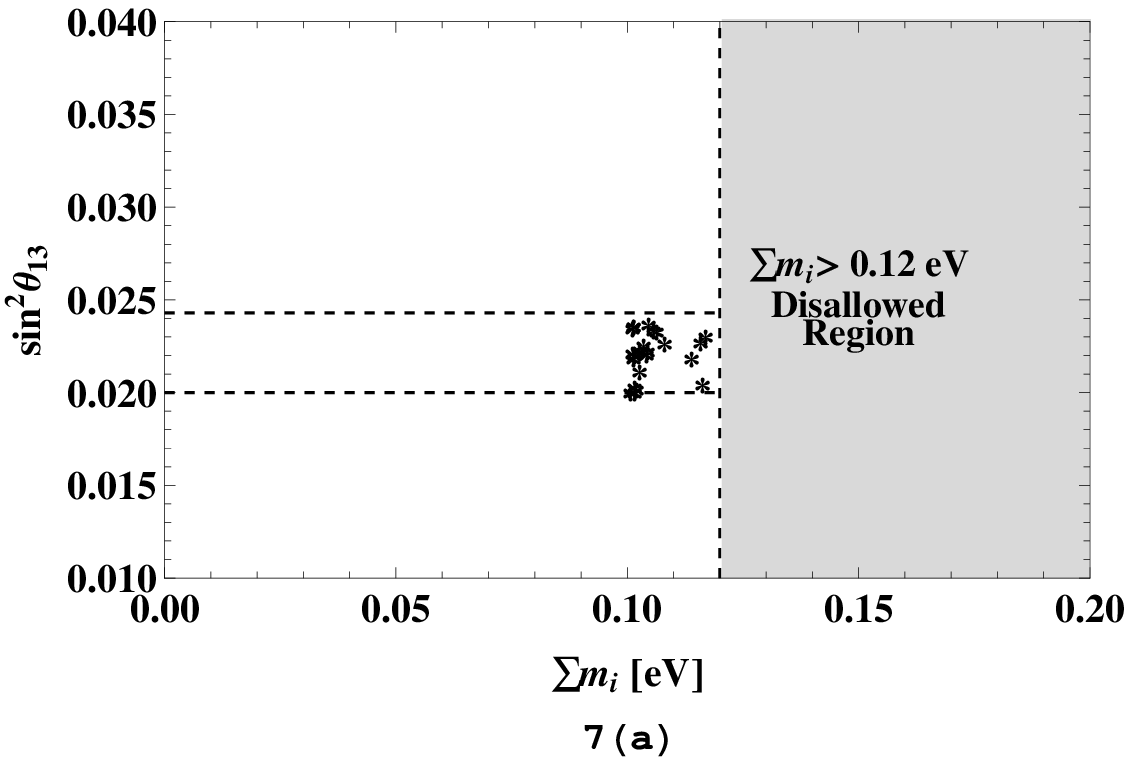,height=6.0cm,width=7.5cm}
				\epsfig{file=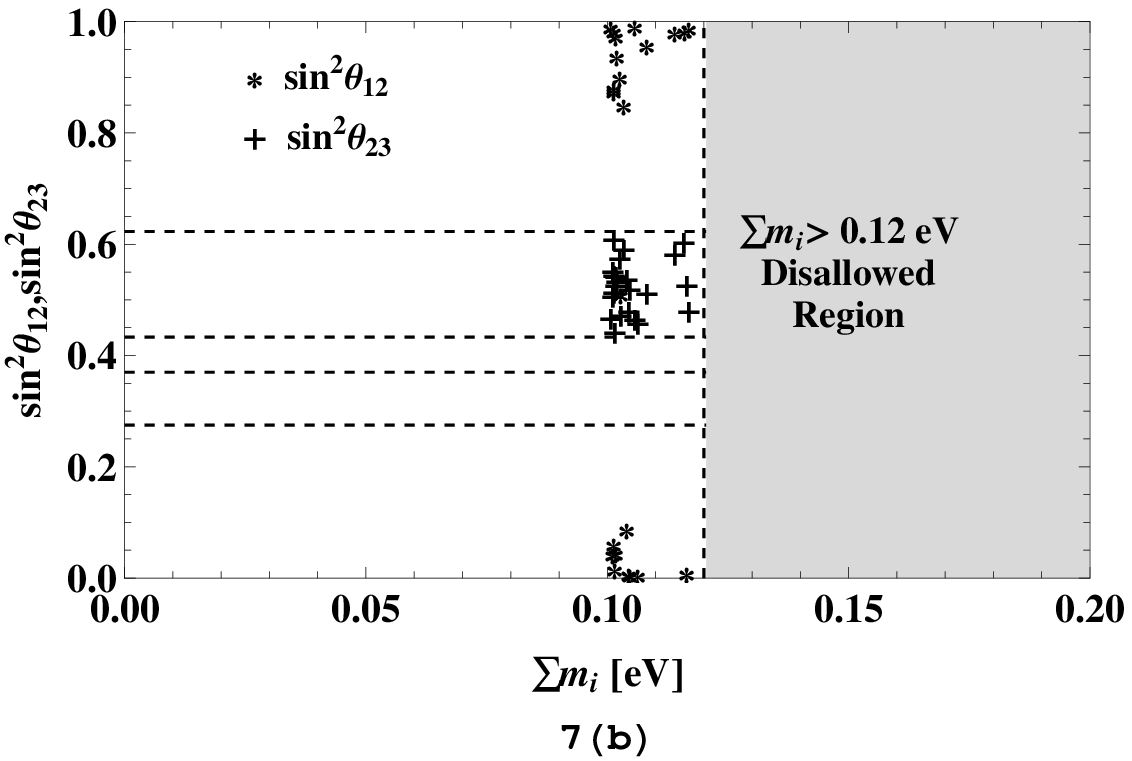,height=6.0cm,width=7.5cm}}
		\end{center}
\caption{\label{fig:7}  The variation of neutrino mixing angles with sum of neutrino masses $\sum m_{i}$  for inverted hierarchy. The grey shaded region is disallowed by cosmological bound on sum of neutrino masses\cite{Giusarma:2016phn,Aghanim:2018eyx}. The horizontal lines represent $3\sigma$ ranges of the mixing angles (Table \ref{tab5}).}
\end{figure}

\noindent In the Fig. \ref{fig:2}, we have depicted the allowed parameter space of complex modulus ($\tau$) in the complex plane. For the allowed parameter space, imaginary part of complex modulus $\tau$ take values greater than one which serves as the source of $CP$ violation. It is evident from Fig. \ref{fig:3}(a) and \ref{fig:3}(b) that the model predicts neutrino mixing angles in consonance with the 3$\sigma$ experimental ranges given in Table \ref{tab5}. The effective Majorana neutrino mass, $M_{ee}$, is an important parameter which could shed the light on the nature of neutrino being Dirac or Majorana particle. In Fig. \ref{fig:4}, we have shown ($\sum m_i$-$M_{ee}$) correlation plot. In fact, the model reproduces the general features of $\sum m_{i}-M_{ee}$ correlation plot in which $M_{ee}$ can be vanishing near the lower bound of $\sum m_{i}$. Also, there is an upward shift in the lower bound on $M_{ee}$ as $\sum m_{i}$ moves towards cosmological upper bound \cite{Giusarma:2016phn,Aghanim:2018eyx}. For example, if $\sum m_{i}= 0.095 eV$ then,  at 3$\sigma$, $M_{ee}>0.012 eV$ which is within the sensitivity reach of $0\nu\beta\beta$ decay experiments such as SuperNEMO \cite{Barabash:2012gc}, KamLAND-Zen \cite{KamLAND-Zen:2016pfg}, NEXT \cite{Granena:2009it,Gomez-Cadenas:2013lta}, nEXO \cite{Licciardi:2017oqg}. Furthermore, the complex modulus $\tau$ is the only source of $CP$-violation which can be estimated in terms of  $CP$-invariants. The Jarlskog $CP$ invariant, $J_{CP}$, related to the Dirac $CP$-phase and $I_{1},I_{2}$ related to the Majorana phases ($\alpha, \beta$) have been defined in Eqns. (\ref{jcp}) and (\ref{i1i2}), respectively. The predictions for these $CP$ invariants are shown in Fig. \ref{fig:5}. It is evident from Fig. \ref{fig:5}(a) and \ref{fig:5}(b) that $|J_{CP}|, |I_{2}| \leq 0.15$ whereas $ |I_{1}| \leq 0.45$. In general, the model predicts existence of both $CP$ conserving and violating solutions. Also, we have plotted the Dirac $CP$-violating phase ($\delta$) with the imaginary part of complex modulus $\tau$ as shown in Fig. \ref{fig:6}. Furthermore, we have scanned the parameter space for inverted hierarchy(IH). We find that model does not satisfy the neutrino oscillation data for IH. In particular, the reactor mixing angle($\sin^{2}\theta_{13}$) (Fig. \ref{fig:7}(a)) and atmospheric mixing angle($\sin^{2}\theta_{23}$) (Fig. \ref{fig:7}(b)) are found to be consistent with $3\sigma$ experimental ranges while $\sin^{2}\theta_{12}$ lies outside the $3\sigma$ experimental range as shown in Fig. \ref{fig:7}(b). \\
\section{Implications for Dark Matter (DM) and Lepton Flavor Violation (LFV)}
The T4-2-$i$ topology generates the neutrino masses at one-loop level with particles running in the loop as potential DM candidates. The scalar potential relevant for inert doublet($\rho$) is
\begin{equation}
    V_{scalar} \supset  \mu_{\rho}^{2}|\rho|^{2} + \lambda_{2}|\rho|^{4}+\lambda_{3}|H|^{2}|\rho|^{2}+\lambda_{4}|H^{\dagger }\rho|^{2}+\lambda_{5}[(H^{\dagger} \rho)^2+h.c.]+\lambda_{6} (\rho^{\dagger}\rho)Tr[\Delta_{\dagger}\Delta]+\lambda_{7}\rho^{\dagger}\Delta\Delta^{\dagger}\rho,
    \end{equation}
where $\lambda_{5}$ transforms as $Y_{6}^{1}$ and $\mu_{\rho}^{2}$, $\lambda_{2,3,4}$ includes the factor $1/(-i \tau+i \bar{\tau})^n$. After spontaneous symmetry breaking, Higgs and scalar triplet field acquire $vevs$ while inert doublet($\rho$) have zero $vev$,
\begin{equation} \label{vev}
	\langle H \rangle=
	\frac{1}{\sqrt{2}}{\begin{pmatrix}
	0 \\ v+h 
		\end{pmatrix}},\hspace{0.2cm}
		\langle \Delta \rangle	=
	\frac{1}{\sqrt{2}}{\begin{pmatrix}
	0 &0\\
    v_{\Delta}+T&0  
		\end{pmatrix}},\hspace{0.2cm}
		\rho	=
	\frac{1}{\sqrt{2}}{\begin{pmatrix}
	\rho_{1}+ i \rho_{2}\\
 \sqrt{2}\rho^{-}
		\end{pmatrix}}.
	\end{equation}

The masses of neutral($\rho_{1}$, $\rho_{2}$) and charged components($\rho^{\pm}$) of inert doublet are given by
\begin{equation}
    M_{\rho_{1},\rho_{2}}^{2}=\mu_{\rho}^{2}+\frac{\lambda_{3}\pm \lambda_{5}}{2}v^{2}+\frac{\lambda_{6}}{2}v_{\Delta}^{2},
\end{equation}
\begin{equation}
    M_{\rho^{\pm}}^{2}=\mu_{\rho}^{2}+\frac{\lambda_{3}+\lambda_{4}}{2}v^{2}+\frac{\lambda_{6}+\lambda_{7}}{2}v_{\Delta}^{2}.
\end{equation}

Here, we have assumed the mixing between neutral components of inert doublets $\rho$ and $\phi$ to be small, thus, neglected in the following analyses of DM and LFV (however, mixing is required for generating non-zero neutrino masses). For simplification, we take masses of $\rho^{\pm}$($\phi^{\pm}$), $\rho_{1}(\phi_1)$, $\rho_{2}(\phi_2$) to be equal to $M_{\rho}(M_{\phi})$. The relic density observed by the Planck collaboration is,
\cite{Planck:2018vyg}
 $$ \Omega h^2 =0.120 \pm 0.001.$$

 \noindent We have explored the scenario of fermionic dark matter where the lightest right-handed neutrino($N_1$) serves as the dark matter candidate. In case, the right-handed neutrino masses are close to each other i.e. $\Delta_{i}=(M_{i}-M_1)/M_{1}<< 1$, the co-annihilation effects are significant leading to DM relic abundance \cite{Jungman:1995df}
 \begin{equation}\label{relic}
 \Omega_{N_1}h^2\approx \frac{3\times 10^{-26} cm^{3}s^{-1}}{\sigma_{eff}}, 
 \end{equation}
 where $\sigma_{eff}$ is the effective cross-section containing co-annihilation effects given by \cite{Ahriche:2017iar}
\begin{equation}\label{sigma}
   \sigma_{eff}= \sum_{i,k=1}^{3} \frac{4}{g_{eff}}(1+\Delta_{i})^{3/2}(1+\Delta_{k})^{3/2}\times e^{-x_{f}(\Delta_{i}+\Delta_{k})}\langle \sigma_{ik}v_{r} \rangle,
\end{equation}
$x_{f}$ is freeze-out temperature and $g_{eff}$ is the effective multiplicity at freeze-out defined as 
\begin{equation}\label{geff}
    g_{eff}(x_{f})=\sum_{i=1}^{3}2(1+\Delta_{i})^{3/2} e^{-x_{f}\Delta_{i}}.
\end{equation} 
The thermally averaged cross-section is given by \cite{Ahriche:2017iar}
\begin{eqnarray}
\nonumber
v_{r}\sigma=&&\frac{|Y_{\rho_{\alpha i}}Y_{\rho_{\beta k}}^*|^{2}}{32\pi}\frac{\sqrt{s^{2}-2s(m^{2}_{l_{\alpha}}+m^{2}_{l_{\beta}})^{2}+(m^{2}_{l_{\alpha}}-m^{2}_{l_{\beta}})^{2}}}{s^{2}-(M^{2}_{i}-M^{2}_{k})^{2}} \bigg[\frac{A_{1}-Q-Q_{2}}{A_{1}}+\frac{Q_{1}Q_{2}}{A_{1}^{2}}\left(1+\frac{B^{2}}{A^{2}_{1}}\right)\\
&& + \delta_{ik} \bigg\{\frac{A_{2}-Q-Q_{2}}{A_{1}}+\frac{Q_{1}Q_{2}}{A_{2}^{2}}\left(1+\frac{B^{2}}{A^{2}_{1}}\right)-\frac{2M_{i}^{2}(s-m^{2}_{l_{\alpha}}-m^{2}_{l_{\beta}})}{A_{2}^{2}}\bigg\} \bigg], 
\end{eqnarray}
where $Y_{\rho_{\alpha i}}$ is couplings of annihilation process of right-handed neutrino into charged leptons ($N_{i}N_{k}\rightarrow l_{\alpha}l_{\beta}^{+}$) mediated by charged component of inert Higgs $\rho$, $ m_{l_{\alpha, \beta}}$ are charged lepton masses and
\begin{eqnarray*}
&&Q_{1}=M_{i}^{2}+m^{2}_{l_{\alpha}}-M^{2}_{\rho^{+}}\hspace{0.4cm} Q_{2}=M_{k}^{2}+m^{2}_{l_{\beta}}-M^{2}_{\rho^{+}},\\
&&A_{1}=\frac{M_{i}^{2}+M^{2}_{k}+m^{2}_{l_{\alpha}}+m^{2}_{l_{\beta}}-2M^{2}_{\rho^{+}}-s}{2}-\frac{\left(M_{i}^{2}-M_{k}^{2}\right)(m^{2}_{l_{\alpha}}-m^{2}_{l_{\beta}})}{2s},\\
&&A_{2}=M_{i}^{2} -M_{\rho}^{2}-\frac{(m^{2}_{l_{\alpha}}+m^{2}_{l_{\beta}}-s)}{2},\\
&&B=2\frac{\sqrt{s^2 -2s(m^{2}_{l_{\alpha}}+m^{2}_{l_{\beta}})+(m^{2}_{l_{\alpha}}-m^{2}_{l_{\beta}})^2}}{2\sqrt{s}}p,
\end{eqnarray*}
$p=\sqrt{s^2 -2s\left(M_{i}^{2}+M_{k}^{2}\right)+\left(M_{i}^{2}-M_{k}^{2}\right)^2}/{2\sqrt{s}}$ is momentum of the incoming right-handed neutrino in the center of mass(COM) frame and $s$ is the total energy of the system in COM frame.
\noindent Also, there is no direct coupling of right-handed neutrino with Higgs boson or $Z$ gauge boson. The effective coupling ($y_{eff}$) at one-loop contributes to spin independent(SI) scattering cross-section \cite{Okada:2013rha}
\begin{equation}\label{sdet}
\sigma_{det}=\frac{y_{eff}^{2}\left(m_{\mathcal{N}}-\frac{7}{9}m_{\mathcal{B}}\right)^{2}m_{\mathcal{N}}^{2} M_{1}^{2}}{4\pi v^{2}m_{h}^{4}\left(m_{\mathcal{N}}+M_{1}\right)^{2}},
\end{equation}
 where $m_{\mathcal{N}}$, $m_{\mathcal{B}}$ are nucleon and baryon masses in the chiral limit. The effective coupling $y_{eff}$ is given by 
 \begin{eqnarray}
\nonumber
y_{eff}=&&\frac{\lambda_{3}v\sum_{\alpha}|Y_{\rho_{\alpha 1}}|^{2}}{16 \pi^{2}M_{1}^{3}}\left(M_{1}^{2}+(M_{\rho^{+}}^{2}-M_{1}^{2}) \ln \left( \frac{M_{\rho^{+}}^{2}-M_{1}^{2}}{M_{\rho^{+}}^{2}}\right) \right) \\
&& -\frac{(\lambda_{3}+\lambda_{4}/2)\sum_{\alpha}|Y_{\rho_{\alpha 1}}|^{2}}{16 \pi^{2}M_{1}^{3}}\left(M_{1}^{2}+(\Bar{m}^{2}-M_{1}^{2}) \ln \left( \frac{\Bar{m}^{2}-M_{1}^{2}}{\Bar{m}^{2}}\right) \right),
\end{eqnarray}
 where $\Bar{m}$ is the average mass of neutral components of inert doublet $\rho$.
\noindent Furthermore, the lepton flavor violation which is highly suppressed in SM can have significant contribution due to new fields introduced by the topology. The most stringent upper bound on LFV branching ratios comes from $\mu\rightarrow e \gamma$ decay \cite{MEG:2013oxv,MEG:2016leq}
 $$Br(\mu\rightarrow e \gamma)< 4.2\times 10^{-13}.$$ 
 \noindent In the model, lepton flavor violation occurs via the mediation of charged and neutral components of inert doublets and right-handed neutrinos which is given by \cite{Ahriche:2022bpx}
 \begin{equation}
Br(\mu\rightarrow e\gamma)=\frac{3\alpha v^{4}}{32\pi }\left|\sum_{i=1}^{3}\sum_{x=\rho,\phi}^{}\frac{Y^{*}_{x_{e i}}Y_{x_{\mu i}}}{M_{x}^{2}}F(M^{2}_{i}/M^{2}_{x^{+}})\right|^2,
\end{equation}
 where $\alpha$ is structure constant, $v$ is vacuum expectation value of the Higgs field. The function $F(x) \left(x=\frac{M^{2}_{i}}{M^{2}_{\rho}}\right)$ is defined as 
 \begin{equation*}
     F(x)=\frac{1-6x+3x^{2}+2x^{3}-6x^{2}\log x}{6(1-x^4)}.
 \end{equation*}

 \begin{figure}[t]
	\begin{center}
			{\epsfig{file=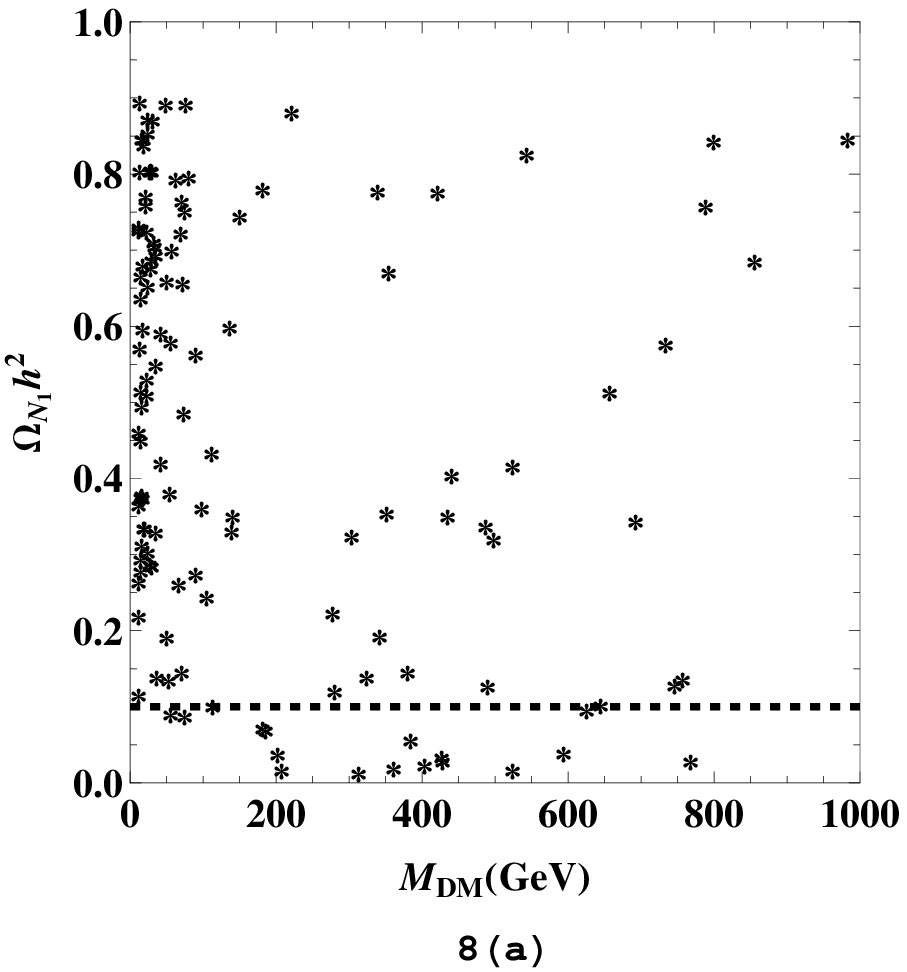,height=6.0cm,width=7.5cm}
				\epsfig{file=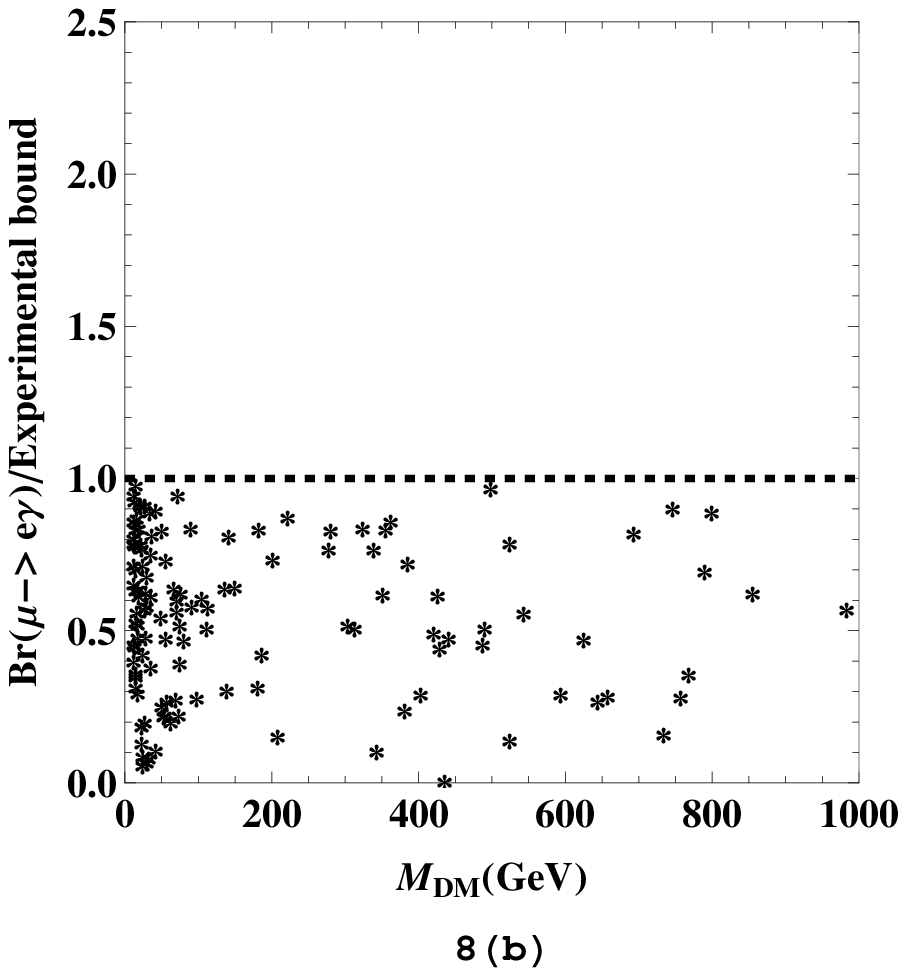,height=6.0cm,width=7.5cm}}
		\end{center}
\caption{\label{fig:8}  The relic density of dark matter and normalized branching ratio of lepton flavor violating (LFV) process $\mu \rightarrow e \gamma$ as a function of dark matter mass ($M_{DM}$). The horizontal lines are the observed relic density of dark matter (Fig. 8(a)) and upper bound on LFV process Br($\mu \rightarrow e \gamma)< 4.2\times10^{-13}$ (Fig. 8(b)), respectively.}
\end{figure}   
\begin{figure}[h]
\begin{center}	
\includegraphics[height=6.0cm,width=7.5cm]{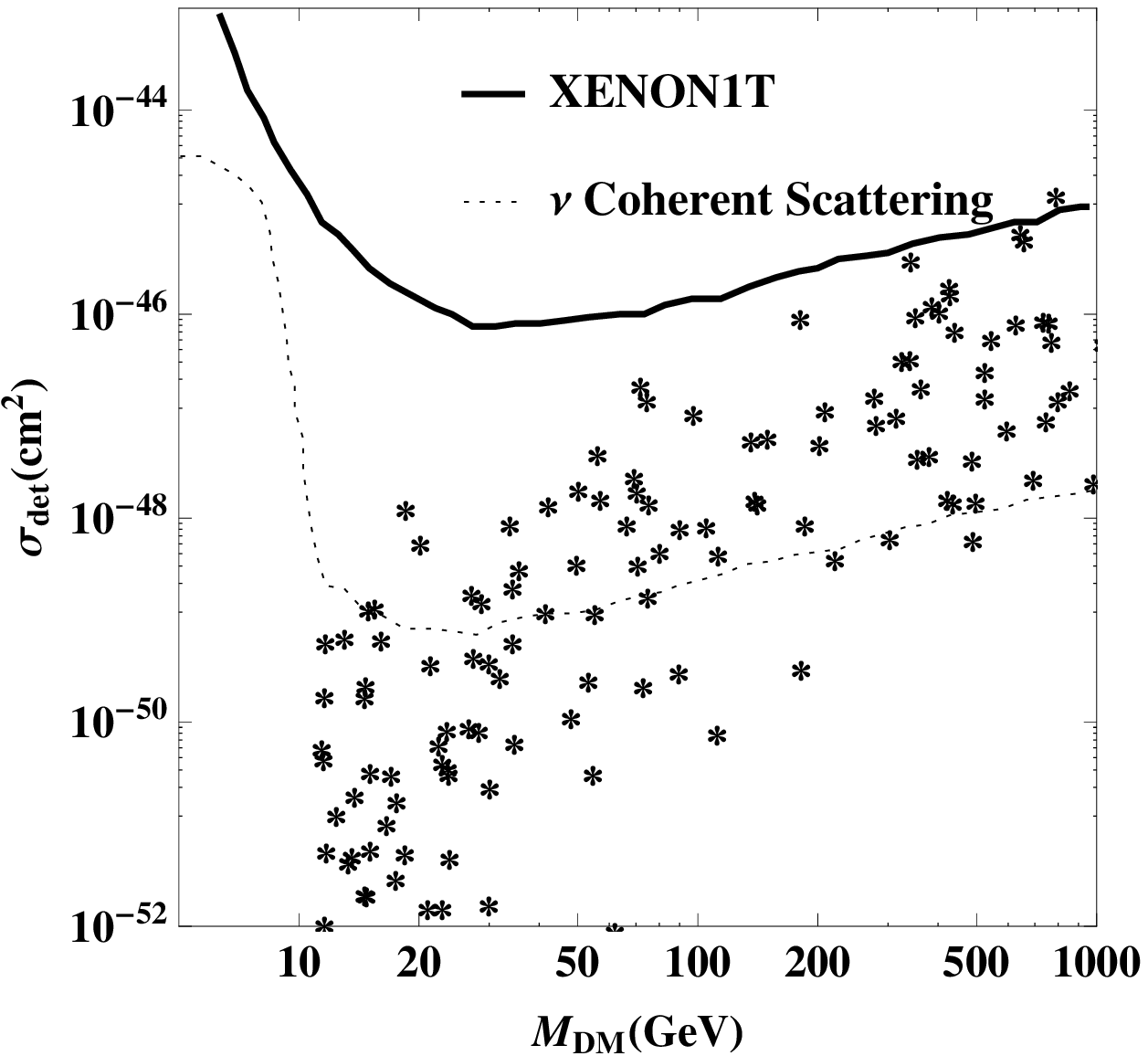}
\end{center}
    \caption{The SI scattering cross-section for direct detection of DM as a function of dark matter mass($M_{DM}$) with the exclusion bound from XENON1T experiment. The dashed line represents the bound of irreducible neutrino background.}
    \label{fig:9}
\end{figure}
\noindent In the numerical analysis, we employed the upper bound on branching ratio of lepton flavor violation process $\mu\rightarrow e\gamma$ and obtain predictions on relic density of DM and SI scattering cross-section for DM direct detection. 

\noindent In Fig. \ref{fig:8}(a), the prediction on relic density of dark matter as a function of dark matter mass($M_{DM}=M_{1}$) is shown. It can be seen that model satisfies the observed relic density in the range $M_{DM}\in (1-10^{3})$ GeV. Also, it is evident from Fig.\ref{fig:8}(b) that model predicts LFV consistent with bound coming from $\mu \rightarrow e \gamma$ process. Further, the implications of the model for dark matter direct detection are shown in Fig. \ref{fig:9}. The points are below the experimental bound of XENON1T\cite{XENON:2018voc,XENON:2019zpr,XENON:2020gfr}. In fact, there is substantial parameter space above neutrino scattering background \cite{Billard:2013qya,Dent:2016iht} which can be probed in the future DM detection experiments. \\

\section{Conclusions}

Out of six topologies of one-loop Weinberg operator at dimension-5, the finite Lorentz structures of T4 topology corresponds to one-loop extension of canonical seesaw scenarios. T4 topology(category (iii)) allows dominant tree-level contribution in addition to one-loop contribution to neutrino mass regardless of imposition of discrete or U(1) symmetry\cite{Bonnet:2012kz}. The discrete or $U(1)$ symmetries have been employed to inhibit tree-level contribution but this requires either enlargement in the field content \cite{Loualidi:2020jlj} or neutrino masses are generated by higher dimensional Weinberg operator\cite{Kanemura:2012rj}. In this work, we propose a possible alternative way to inhibit tree-level contribution wherein one may not require additional fields (fields other than required by the topology) provided we work within the paradigm of modular symmetry. Specifically, we have constructed a possible realization of T4-2-$i$ topological Lorentz structure based on A$_4$ modular symmetry wherein neutrino masses are generated through one-loop dimension-5 Weinberg operator. The odd  modular weight of loop particles \textit{viz.} right-handed Majorana neutrinos ($N_{i}$), inert doublets ($\rho, \phi$) ensures the dark matter stability. The model predictions for neutrino mixing angles and mass-squared differences are found to be in consonance with the current neutrino oscillation data. We have, also, obtained the implication on effective Majorana mass $M_{ee}$ and $CP$ invariants as shown in Figs. \ref{fig:4} and \ref{fig:5}, respectively. In fact, for $\sum m_{i}= 0.10 $ eV, the effective Majorana mass $M_{ee}>0.012 $ eV, at 3$\sigma$, which is within the sensitivity reach of $0\nu\beta\beta$ decay experiments. The model, in general, is consistent with both $CP$ conserving and violating solutions. In this work, we have, also, explored the possibility of fermionic dark matter where one of the right-handed neutrino ($N_1$) is assumed to be lightest. We have obtained the predictions of the model for relic density of DM and its implication for direct searches experiments while obeying the bound from lepton number violating process $\mu\rightarrow e \gamma$. We find that the model explains observed relic density for the DM mass in the range ($1-10^{3}$) GeV. Furthermore, SI scattering cross-section is 
 found to be below the experimental bound of XENON1T and substantial parameter space above neutrino scattering background is viable which can be probed in the future DM direct detection experiments.

\begin{appendices}
\section{Modular symmetry and Permutation groups}\label{appA}

For a complex number $\tau$, the linear fractional transformations are defined as 
\begin{equation}\label{A1}
\gamma : \tau \rightarrow\gamma(\tau) =\frac{a\tau +b}{c\tau+d}, 
\end{equation}
where $a$, $b$, $c$ and $d$ are integers($Z$) obeying the condition $ad-bc=1$. These linear fractional transformations acting on the upper-half of the complex plane constitute modular group ($\Gamma$). The generators of the modular group satisfy the relations $S^2=I$ and $(ST)^3=I$ and are defined by matrices
\[ S= \left( \begin{array}{cc}
0 & 1 \\
-1 & 0
\end{array} \right);
\hspace{0.7cm}
T=\left( \begin{array}{cc}
1 & 1 \\
0 & 1
\end{array} \right). 
\]
The action of these generators on the complex plane is defined as
$$ S: \tau \rightarrow -\frac{1}{\tau},\hspace{0.7 cm} T: \tau \rightarrow \tau+1.$$
Also, $\Gamma$(N) is defined as series of groups given by
\begin{equation}\label{A2}
\Gamma(N)=\left\{
 \begin{bmatrix}
    a & b\\
     c &d
  \end{bmatrix} \hspace{0.2cm}
  \in \hspace{0.2cm}  SL(2,Z),\hspace{0.2cm}\begin{bmatrix}
    a & b\\
     c &d
  \end{bmatrix}=\begin{bmatrix}
    1 & 0\\
     0 &1
  \end{bmatrix} (mod N) 
  \right\},    
\end{equation}
where, $SL(2,Z)$ is a special linear group of $2\times2$ matrices with unit determinant and $\Gamma(1)=SL(2,Z)$. In general, $\gamma$ and $-\gamma$ result in identical linear transformations  however, distinct linear transformations form series of infinite modular groups $\Bar{\Gamma}(N)$ such that $\Bar{\Gamma}=\Bar{\Gamma}(1)=\Gamma(1)/\{I,-I\}=PSL(2,Z)$. The quotient group of infinite modular groups defined as  $\Gamma_N$ = $\Bar{\Gamma}/\Bar{\Gamma}(N)$ leads to finite Modular groups. These groups ($\Gamma_N$) are isomorphic to the permutation groups such as $\Gamma_{2} \simeq S_{3}$ \cite{Kobayashi:2018vbk,Okada:2019xqk}, $\Gamma_{3} \simeq A_{4}$ \cite{Nomura:2019jxj,Zhang:2019ngf,Okada:2021aoi,Behera:2020sfe,Behera:2022wco}, $\Gamma_{4} \simeq S_{4}$ \cite{King:2019vhv,Penedo:2018nmg,Kobayashi:2019mna} and $\Gamma_{5} \simeq A_{5}$ \cite{Novichkov:2018nkm,Ding:2019xna}.\\

\noindent In general, fractional linear transformations are very constraining and are not trivial to be exactly preserved, however, one may define the modular forms, $f(\tau$) as holomorphic functions of complex modulus $\tau$ such that fractional linear transformations preserve the zeroes and poles of $f(\tau)$. Under the transformation properties of $\Gamma(N)$, described in Eqns. (\ref{A1}) and (\ref{A2}), level $N$ modular forms are defined as
\begin{equation}\label{A3}
  f(\gamma \tau)= (c\tau + d)^{2k} f(\tau),  
\end{equation}
where
 $$  \hspace{0.6cm} \gamma= \begin{bmatrix}
    a & b\\
     c & d
  \end{bmatrix}
  \hspace{0.2cm}\in \hspace{0.4cm} \Gamma(N),
$$
up to the factor $(c\tau + d)^{2k}$ with modular weight $2k$. In general, modular forms are invariant under $\Gamma(N)$, however, they do transform under finite modular group $\Gamma_N$. In fact, It result in a unique feature of the finite modular groups that Yukawa couplings transform under the modular symmetry.
\noindent In unitary representation, the modular transformations can be written as 
\begin{equation}\label{A4}
   f(\tau)\rightarrow e^{i \alpha} (c \tau +d)^{k} f(\tau).   
  \end{equation}  
  In particular, for modular weight 2, Eqn. (\ref{A4}) result in
  $$ \frac{d}{d\tau}\log f(\tau) \rightarrow (c\tau +d)^2 \frac{d}{d\tau}\log f(\tau) + kc (c\tau+d), $$
  such that the last term $kc (c\tau+d)$  should vanish. Similarly, it should, also, vanish for all higher modular weights to preserve the fractional linear transformations. This leads to the constraint $\sum  k_{i} =0$ if $\Gamma_N$ modular symmetry is to be preserved.
 \noindent The Dedekind $\eta$-function and its derivative can be used to construct modular forms of weight 2 as \cite{Feruglio:2017spp} 
    \begin{equation}
        \begin{aligned}
  Y_{1}^{2}(\tau) &= \frac{i}{2 \pi} \left[ 
 \frac{\eta^{'}(\tau/3)}{\eta(\tau/3)} + \frac{\eta^{'}((\tau+1)/3)}{\eta((\tau+1)/3)}+\frac{\eta^{'}((\tau+2)/3)}{\eta((\tau+2)/3)}-27\frac{\eta^{'}(3\tau)}{\eta(3\tau)}\right],\\
  Y_{2}^{2}(\tau) &= \frac{-i}{\pi} \left[ 
 \frac{\eta^{'}(\tau/3)}{\eta(\tau/3)} + \omega^{2}\frac{\eta^{'}((\tau+1)/3)}{\eta((\tau+1)/3)}+ \omega \frac{\eta^{'}((\tau+2)/3)}{\eta((\tau+2)/3)}\right],\\
 Y_{3}^{2}(\tau) &= \frac{-i}{\pi} \left[ 
 \frac{\eta^{'}(\tau/3)}{\eta(\tau/3)} + \omega \frac{\eta^{'}((\tau+1)/3)}{\eta((\tau+1)/3)}+ \omega^{2} \frac{\eta^{'}((\tau+2)/3)}{\eta((\tau+2)/3)}\right],
\end{aligned}
    \end{equation}  
where $\omega=e^{i2\pi/3}$ and 
 $
   Y_{2}^{2}+2 Y_{1}^{2} Y_{3}^{2} =0$.
   
\noindent These modular forms of weight 2 can be arranged as $A_4$ triplet \textit{viz.} 
 $$ Y= \begin{pmatrix}
	Y_{1}^{2} \\
	Y_{2}^{2} \\
	Y_{3}^{2} \\
	\end{pmatrix}.$$
	Also, modular forms of higher weights ($2,4,6,8.....$) can be constructed using modular forms of weight 2. For example, the Yukawa couplings of modular weight 4 consists of two singlets $1$, $1'$ and one triplet $3$ as 
\[
Y_{1}^{4}=\left((Y_{1}^{2})^2+2 Y_{2}^{2} Y_{3}^{2}\right), \quad Y_{1^{\prime}}^{4}=\left((Y_{3}^{2})^2+2 Y_{1}^{2} Y_{2}^{2}\right), \quad Y_{3}^{4}=
\left(\begin{array}{l}
(Y_{1}^{2})^2-Y_{2}^{2} Y_{3}^{2} \\
(Y_{3}^{2})^2-Y_{1}^{2} Y_{2}^{2} \\
(Y_{2}^{2})^2-Y_{1}^{2} Y_{3}^{2}
\end{array}\right).
\]
It is to be noted that, in numerical analysis, we have used the $q$-expansion of modular forms using Dedekind $\eta$-function defined in upper-half of the complex plane as
      \begin{equation}
          \eta(\tau)=q^{1/24} \sum_{n=1}^{\infty} (1-q^{n}), 
      \end{equation}
where $q=e^{i2\pi \tau}$. 
For the Yukawa couplings of modular weight 2, the $q$-expansions of Dedekind eta-function is
\begin{equation}
\begin{aligned}
&Y_{1}^{2}(\tau)=1+12 q+36 q^{2}+12 q^{3}+\ldots ,\\
&Y_{2}^{2}(\tau)=-6 q^{1 / 3}\left(1+7 q+8 q^{2}+\ldots\right), \\
&Y_{3}^{2}(\tau)=-18 q^{2 / 3}\left(1+2 q+5 q^{2}+\ldots\right).
\end{aligned}
\end{equation}
\section{Supersymmetric Scalar Potential}\label{appB} 
The F-terms are given by
\begin{eqnarray*}
    &&|F_{u}|^{2}=|\mu_{1}^{2} H_{d}+\lambda_{3} \Delta_{u}H_{u}|,\\
    &&|F_{d}|^{2}=|\mu_{1}^{2} H_{u}+\lambda_{u} \Delta_{d}H_{d}|,\\
    &&|F_{\Delta_{u}}|^{2}=|\mu_{\Delta} \Delta_{d}+\lambda_{3} H_{u}H_{d}+\lambda_{6} \rho_{D}\phi{u}|,\\
    &&|F_{\Delta_{d}}|^{2}=|\mu_{\Delta} \Delta_{u}+\lambda_{4} H_{d}H_{d}+\lambda_{5} \rho_{u}\phi_{d}|,\\
&&|F_{\rho_{\mu}}|^{2}=\left|\mu_{3}^{2}\rho_{d}+\mu_{4}^{2}\Phi_{u}+\lambda_{5}\Delta_{d}\Phi_{d}\right|,\\
     &&|F_{\rho_{d}}|^{2}=\left|\mu_{3}^{2}\rho_{u}+\mu_{5}^{2}\Phi_{d}+\lambda_{6}\Delta_{u}\Phi_{u}\right|,\\
&&|F_{\Phi_{\mu}}|^{2}=\left|\mu_{2}^{2}\Phi_{d}+\mu_{4}^{2}\rho_{u}+\lambda_{6}\rho_{d}\Delta_{u}\right|, \\
    &&|F_{\Phi_{d}}|^{2}=\left|\mu_{2}^{2}\Phi_{u}+\mu_{5}^{2}\rho_{d}+\lambda_{5}\rho_{u}\Delta_{d}\right|,
\end{eqnarray*}
\noindent while D-terms are
\begin{equation*}
  D^{2}=\dfrac{g_{1}^{2}}{2}\left[\frac{1}{2}\left(H_{u}^{\dagger}H_{u}-H_{d}^{\dagger}H_{d}+Tr(\Delta_{d}^{\dagger}\Delta_{d})-Tr(\Delta_{u}^{\dagger}\Delta_{u})\right)\right]^{2} , 
\end{equation*}
and
\begin{equation*}
  \vec{D}^{2}=\dfrac{g_{2}^{2}}{2} \sum_{a=1}^{3} \left[\frac{1}{2}\left(H_{u}^{\dagger}\sigma^{a} H_{u}+H_{d}^{\dagger}\sigma^{a}H_{d}+\frac{1}{2}Tr(\Delta_{d}^{\dagger}[\sigma^{a},\Delta_{d}])+\frac{1}{2}Tr(\Delta_{\mu}^{\dagger}[\sigma^{a},\Delta_{u}])\right)\right]^{2}. 
\end{equation*}

\end{appendices}

\section*{Acknowledgments}
 M. K. acknowledges the financial support provided by Department of Science and Technology(DST), Government of India vide Grant No. DST/INSPIRE Fellowship/2018/IF180327. The authors, also, acknowledge Department of Physics and Astronomical Science for providing necessary facility to carry out this work.

% The bibliography will probably be heavily edited during typesetting.
% We'll parse it and, using the arxiv number or the journal data, will
% query inspire, trying to verify the data (this will probalby spot
% eventual typos) and retrive the document DOI and eventual errata.
% We however suggest to always provide author, title and journal data:
% in short all the informations that clearly identify a document.

\end{document}